\documentclass[lettersize,journal]{IEEEtran}
\usepackage{amsmath,amsfonts}
\usepackage{array}
\usepackage[caption=false,font=normalsize,labelfont=sf,textfont=sf]{subfig}
\usepackage{textcomp}
\usepackage{stfloats}
\usepackage{url}
\usepackage{verbatim}
\usepackage{graphicx}
\usepackage{cite}
\usepackage{amssymb}
\usepackage{setspace}
\usepackage{epstopdf}
\usepackage{indentfirst}
\usepackage{svg}
\usepackage{caption}
\usepackage{textcomp}
\usepackage{float}
\allowdisplaybreaks[4]
\usepackage{graphicx}
\usepackage{epsfig}
\usepackage{array}
\usepackage{mathrsfs}
\usepackage{enumitem}
\usepackage{hyperref}
\hypersetup{colorlinks=true,
	linkcolor=teal,
	anchorcolor=blue,
	citecolor=blue,
	urlcolor=black
}
\usepackage{color}
\usepackage{longtable}
\usepackage{epsfig}
\usepackage{mathrsfs}
\usepackage{tipa}
\usepackage{booktabs}
\usepackage{caption}
\usepackage{multirow}
\usepackage{algorithm}
\usepackage{algorithmic}
\usepackage{epstopdf}

\def\BibTeX{{\rm B\kern-.05em{\sc i\kern-.025em b}\kern-.08em
		T\kern-.1667em\lower.7ex\hbox{E}\kern-.125emX}}
\usepackage{enumitem}
\setlist[itemize]{leftmargin=*}

\usepackage{etoolbox}
\AtBeginEnvironment{algorithm}{\setlength{\intextsep}{5pt}} 
\begin{document}
	\setstretch{0.957}
	\title{\textcolor{black}{Minimum-Cost State-Flipped} Control\\ for Reachability of Boolean Control Networks \\using Reinforcement Learning}
	
	\author{Jingjie Ni, Yang Tang,~\IEEEmembership{Fellow,~IEEE,} Fangfei Li,~\IEEEmembership{Member,~IEEE}
		\thanks{This work is supported by  the National Natural Science Foundation of China under Grants 62173142, 62233005, the Programme of Introducing Talents of Discipline to Universities (the 111 Project) under Grant B17017, and the Fundamental Research Funds for the Central Universities under Grant JKM01231838.}
		\thanks{Jingjie Ni is with the School of Mathematics, East China University
			of Science and Technology, Shanghai, 200237, P.R. China (email: nijingjie20000504@163.com).}
		\thanks{Yang Tang is with the Key Laboratory of Smart Manufacturing in Energy
			Chemical Process, Ministry of Education, East China University of Science
			and Technology, Shanghai, 200237, P.R. China (email: yangtang@ecust.edu.cn).}
		\thanks{Fangfei Li is with the School of Mathematics \textcolor{black}{and the Key Laboratory of Smart Manufacturing in Energy
				Chemical Process, Ministry of Education,} East China University
			of Science and Technology, Shanghai, 200237, P.R. China (email: li\_fangfei@163.com, lifangfei@ecust.edu.cn).}
	}

	\markboth{Journal of \LaTeX\ Class Files,~Vol.~14, No.~8, August~2021}%
	{Shell \MakeLowercase{\textit{et al.}}: A Sample Article Using IEEEtran.cls for IEEE Journals}

	\maketitle
	
	\begin{abstract}
		This paper proposes \textsl{model-free} reinforcement learning methods for minimum-cost state-flipped control in Boolean control networks (BCNs). We tackle two questions: 1) finding the flipping kernel, namely the flip set with the smallest cardinality ensuring reachability, and 2) deriving optimal policies to minimize the number of flipping actions for reachability based on the obtained flipping kernel.
		For question 1), \textsl{Q}-learning's capability in determining reachability is demonstrated. To expedite convergence, we incorporate two improvements: i) demonstrating that previously reachable states remain reachable after adding elements to the flip set, followed by employing transfer learning, and ii) \textcolor{black}{initiating each episode with special initial states whose reachability to the target state set are currently unknown.}
		Question 2) requires optimal control with terminal constraints, while \textsl{Q}-learning only handles unconstrained problems. To bridge the gap, we propose a BCN-characteristics-based reward scheme and prove its optimality.
		Questions 1) and 2) with large-scale BCNs are addressed by employing small memory \textsl{Q}-learning, which reduces memory usage by only recording visited action-values. An upper bound on memory usage is provided to assess the algorithm's feasibility. 
		\textcolor{black}{To expedite convergence for question 2) in large-scale BCNs, we introduce adaptive variable rewards based on the known maximum steps needed to reach the target state set without cycles.}
		Finally, the effectiveness of the proposed methods is validated on both small- and large-scale BCNs.

	\end{abstract}
	
	
	\vspace{-5pt}
	\section{Introduction}
	Gene regulatory networks are a crucial focus in systems biology. The concept of Boolean networks was first introduced in 1969 by \textcolor{black}{Kauffman} \cite{BN1969} as the earliest model for gene regulatory networks. Boolean networks consist of several Boolean variables, each of which takes a value of ``0” or ``1” to signify whether the gene is transcribed or not. As gene regulatory networks often involve external inputs, Boolean networks were further developed into Boolean control networks (BCNs), which incorporate control inputs to better describe network behaviors.
	
	State-flipped control is an innovative method used to change the state of specific nodes in BCNs, flipping them from ``1" to ``0" or from ``0" to ``1". In the context of systems biology, state-flipped control can be achieved through gene regulation techniques such as transcription activator-like effector repression and clustered regularly interspaced short palindromic repeats activation \cite{boettcher_choosing_2015}. \textcolor{black}{State-flipped control is a powerful control method that minimally disrupts the system structure.} \textcolor{black}{From a control effectiveness standpoint, state-flipped control enables BCNs to achieve any desired state with an in-degree greater than 0 \cite{flip8,flip6}. In contrast, state feedback control can only steer BCNs towards the original reachable state set, which is a subset of states with an in-degree greater than 0. Another highly effective control method is pinning control, but it has the drawback of causing damage to the network structure, unlike state-flipped control.
	State-flipped control was first proposed by Rafimanzelat to study the controllability of attractors in BNs \cite{flip1}. Subsequently, to achieve stabilization, authors in \cite{flip2} and \cite{flip5} respectively considered flipping a subset of nodes after the BN had passed its transient period and flipping a subset of nodes in BCNs at the initial step. Zhang et al. further extended the concept of state-flipped control from BCNs to switched BCNs \cite{flip3}. Considering that previous research has primarily focused on one-time flipping \cite{flip1, flip2, flip3, flip5}, which may hinder the achievement of certain control objectives, a more comprehensive form of state-flipped control that permits multiple flipping actions was introduced to study stabilization in BNs and BCNs \cite{flip8, flip6}.} Furthermore, to minimize control costs, the concept of flipping kernel was proposed \cite{flip6, flip8}, representing the smallest set of nodes that can accomplish the control objectives. In this paper, we extend the existing studies by investigating problems including finding flipping kernels, under joint control proposed by \cite{flip8}. These joint controls involve the combination of state-flipped controls and control inputs \textcolor{black}{as defined in \cite{flip8}}.
	
	Closely tied to stabilization and controllability, reachability is a prominent area that requires extensive exploration. It involves determining whether a system can transition from an initial subset of states to a desired subset of states. This concept holds significant importance in domains such as genetic reprogramming and targeted drug development, where the ability to transform networks from unfavorable states to desirable ones is pivotal \cite{flip2}. Previous studies \cite{reach1, reach2, reach3, reach4} have proposed various necessary and sufficient conditions for reachability in BCNs and their extensions. Additionally, an algorithm has been developed to identify an optimal control policy that achieves reachability in the shortest possible time \cite{reach2}. In the context of BCNs under state-flipped control, the analysis of reachability between two states can be conducted using the semi-tensor product \cite{flip8}.
	
	Despite extensive studies on the reachability of BCNs, there are still unresolved issues in this field. These problems can be categorized into three aspects. 
	\renewcommand{\labelenumi}{\noindent{\alph*.}}
	\begin{enumerate}[itemsep=0pt,topsep=0pt,parsep=0pt, leftmargin=12pt]
		\item Firstly, the existing literature \cite{reach1, reach2, reach3, reach4} focuses on determining whether reachability can be achieved under certain control policies, without optimizing control costs. This is significant in both practical applications and theoretical research. For instance, in the development of targeted drugs, an increasing number of target points raises the difficulty level. Additionally, it is desirable to minimize the frequency of drug usage by patients to achieve expense savings and reduce drug side effects. If we formulate this problem as an optimal control problem, our objective is to find the flipping kernel and minimize the number of flipping actions. \textcolor{black}{Emphasizing the identification of flipping kernels is more critical than minimizing flipping actions, as it significantly reduces the dimensionality of joint control, thereby exponentially alleviating the computational complexity when minimizing flipping actions. It is worth mentioning that achieving reachability while optimizing control costs poses an optimization problem with terminal constraint.} \textcolor{black}{To the best of our knowledge, conventional optimal control techniques such as the path integral approach and policy iteration \cite{opts2022,PI2019WU, KLsamplebased, finite2018zhu}, which rely on predetermined cost functions, are deemed unsuitable for our study. This is primarily due to the challenge of expressing the objective of simultaneously minimizing control costs and achieving terminal goals, namely, ``achieving reachability," solely through a cost function.} 
		\item \textcolor{black}{Secondly, the model-free scenario requires special consideration, especially considering the increasing complexity associated with accurately modeling systems with numerous BCN nodes. While model-free approaches exist in the PBCNs field \cite{QLStabilityPBCN2021, QLForest, flip6, flip8, flip9, pxl}, they address different challenges compared to ours. Additionally, there are ongoing investigations into cost reduction problems \cite{reach1, reach2, reach3, reach4, mintime, MinimumN2021, MinimumC2022}. However, these studies are conducted under the assumption of known or partially known models.}
		\item Thirdly, the \textcolor{black}{existing methods in the fields of BCN reachability\cite{reach1, reach2, reach3, reach4, mintime} are limited in their applicability to small-scale BCNs. Several interesting approaches have been proposed for handling large BCNs, emphasizing controllability and stability \cite{zhu2023minimal, zhu2023towards, zhu2022strong}. However, these references concentrate on optimization goals slightly different from ours. They aim to minimize the set of controlled nodes, namely, the dimension of the controller. In contrast, our objective is not only to reduce the controller dimension but also to further minimize flipping actions, namely, the frequency of control implementation.}
	\end{enumerate}
	
	Considering the aforementioned issues, we aim to design minimum-cost state-flipped control for achieving reachability in BCNs, in the absence of a known system model. Specifically, we address the following questions:
	\renewcommand{\labelenumi}{{ \arabic*)}}
	\begin{enumerate}[itemsep=1pt,topsep=1pt,parsep=1pt, leftmargin=*]
		\item How to find the flipping kernel for reachability?
		\item Building on question 1), how can we determine the control policies that minimize the number of required flipping actions to achieve reachability?
	\end{enumerate}
	Questions 1) and 2) are the same as 1) and 2) mentioned in the abstract. They are reiterated here for clarity purposes.
	
	To tackle these questions, we turn to reinforcement learning-based methods. Compared with the semi-tensor product approach, reinforcement learning-based methods are suitable for model-free scenarios, which aligns with our considered setting. Moreover, reinforcement learning eliminates the need for matrix calculations and offers the potential to handle large-scale BCNs. 
	In particular, we consider the extensively applied $Q$-learning ($Q$L) method, which has been successful in solving control problems in BCNs and their extensions \cite{QLStabilityPBCN2021, QLForest, flip6, flip8, flip9}. 
	For reachability problems of large-scale BCNs, an enhanced $Q$L method has been proposed by \cite{pxl}, providing a reliable foundation for this study. When compared to the deep $Q$-learning method \cite{2016DDQN}, which is also suitable for large-scale BCNs, \textcolor{black}{the method proposed in \cite{pxl}} is favored due to its theoretical guarantees of convergence \cite{QLcovergence}.

	Taken together, we propose an improved $Q$L to address questions 1)- 3). Our work's main contributions are as follows:
	\renewcommand{\labelenumi}{{\arabic*.}}
	\begin{enumerate}[itemsep=0pt, topsep=0pt, parsep=0pt, leftmargin=*]
		\item Regarding question 1), unlike \textcolor{black}{the semi-tensor product method proposed in} \cite{flip6, flip8}, \textcolor{black}{our approach is model-free}. We demonstrate that reachability can be determined using $Q$L and propose improved $Q$L with faster convergence \textcolor{black}{compared to the traditional one \cite{QL}}. Firstly, transfer learning (TL) \cite{transfer} is integrated under the premise that the transferred knowledge is proven to be effective. Then, special initial states are incorporated into $Q$L, allowing each episode to begin with an unknown reachability state, thereby eliminating unnecessary computations.
		\item Question 2) involves optimal control with terminal constraints, while $Q$L only handles unconstrained problems. To bridge this gap, a BCN-characteristics-based reward scheme is proposed. \textcolor{black}{Compared to other model-free approaches \cite{QLStabilityPBCN2021, QLForest, flip6, flip8, flip9,pxl}, we \textcolor{black}{first} tackle a complex problem while providing rigorous theoretical guarantees.}
		\item Compared to matrix-based methods \cite{flip6, flip8} and \textcolor{black}{deep reinforcement} techniques \cite{QLStabilityPBCN2021, flip9, QLForest}, our algorithms are suitable for large-scale BCNs \textcolor{black}{with convergence guarantee}. \textcolor{black}{In contrast to \cite{pxl}, the} memory usage in the presented problem is clarified to be not directly related to the scale of the BCNs, and an upper bound is provided to evaluate algorithm suitability. \textcolor{black}{Further,} To accelerate convergence in question 2), a modified reward scheme is introduced.
	\end{enumerate}
	
	\textcolor{black}{
		The paper progresses as follows: 
		Section II provides an overview of the system model and defines the problem statement. Section III introduces the reinforcement learning method. In Section IV, fast iterative  $Q$L and small memory iterative $Q$L for large-scale BCNs are presented to identify the flipping kernel for reachability. In Section V, to minimize the number of flipping actions, $Q$L with a BCN-characteristics-based reward setting and that with dynamic rewards for large-scale BCNs are proposed. Algorithm complexity is detailed in Section VI. Section VII validates the proposed method. Finally, conclusions are drawn in Section VIII. 
	}
	
	We use the following notations throughout this paper:
	\begin{itemize}[itemsep=0.85pt,topsep=0.85pt,parsep=0.85pt]
		\item $\mathbb{Z^+}$, $\mathbb{R}^{+} $, $\mathbb{R}$, $\mathbb{R}^{m\times n}$ denote the sets of non-negative integers, non-negative real numbers, real numbers, and \textcolor{black}{${m\times n}$ real number matrices,} respectively.
		\item $\mathbb{E}[\cdot] $ denotes the expected value operator and var$[\cdot] $ denotes the variance. $ \Pr\left\{E_1\mid E_2\right\} $ denotes the probability of event $E_1$ occurring given that event $E_2$ has occurred. 
		\item For a set $S$, $|S|$ denotes \textcolor{black}{the cardinal number of} $S$. 
		\item For $M_{m\times n}\subset\mathbb{R}^{m\times n}$, $|M_{m\times n}|_\infty$ denotes the maximal value of the elements in $M_{m\times n}$. 
		\item The relative complement of set $S_2$ in set $S_1$ is denoted as $S_1\setminus S_2$.
		\item There are four basic operations on Boolean variables, which are ``not", ``and", ``or", and ``exclusive or", expressed as $\neg$, $\wedge$, $\vee$, and $\oplus$, respectively.
		\item $\mathcal{B}:=\{0,1\}$, and $\mathcal{B}^{n}:=\underbrace{\mathcal{B} \times \ldots \times \mathcal{B}}_{n}$.
	\end{itemize}
	
	\vspace{-5pt}
	\section{Problem Formulation}\label{sec:section2}
	\textcolor{black}{This section introduces the system model, with a specific focus on BCNs under state-flipped control. Subsequently, we outline the problems under investigation.}
	
	\vspace{-5pt}
	\subsection{System Model}
	\subsubsection{BCNs}A BCN with $n$ nodes and $m$ control inputs is defined as
	\begin{equation}
		\begin{aligned}
			x_{i}(t+1)=f_{i}\Big(x_{1}(t),..., x_{n}(t),& u_{1}(t),..., u_{m}(t)\Big), \\
			& t\in \mathbb{Z^{+}},  \ i=1,...,n,
		\end{aligned}
		\label{eqBCN}
	\end{equation}
	where $f_{i}: \mathcal{B}^{n+m} \rightarrow \mathcal{B} , i\in \{1,...,n\}$ is a logical function, $x_i(t)\in \mathcal{B}, i \in \{1,...,n\}$ represents the $i^{th}$ node at time step $t$, and $u_j(t)\in \mathcal{B}, j\in \{1,...,m\}$ represents the $j^{th}$ control input at time step $t$. All nodes at time step $t$ are grouped together in the vector $x(t)=\left(x_{1}(t), \ldots, x_{n}(t) \right) \in \mathcal{B}^{n} $. Similarly, all control inputs at time step $t$ are represented by $u(t)=\left(u_{1}(t), \ldots, u_{m}(t) \right) \in \mathcal{B}^{m} $. 
	
	\subsubsection{State-flipped Control}
	Considering that not all nodes in a BCN can be flipped, we refer to the set of all flippable nodes as a flip set $A\subseteq \{1,2,...,n\}$. At each time step $t$, a flipping action $A(t) = \{i_{1}, i_{2},..., i_{k}\} \subseteq A$ is selected. According to the flipping action $A(t)$, the flip function is defined as
	\begin{equation}
		\begin{aligned}
			\eta_{A(t)}\Big(x(t)\Big)= 
			&\left. \Big(x_1(t),..., \neg x_{i_{1}}(t),..., \right.\\
			&\left. \neg x_{i_{2}}(t), ...,\neg x_{i_{k}}(t),..., x_n(t)\Big). \right.
		\end{aligned}
		\label{eqflipcontrol}
	\end{equation}
	Note that to achieve a specific control objective, it is not necessary to flip all the nodes in the set $A$. The flipping kernel, denoted as $B^*\subset A$, is defined as the flip set with the minimum cardinality required to achieve reachability.
	
	\subsubsection{BCNs under State-flipped Control}Based on the definition of BCNs and state-flipped control, a BCN with $n$ nodes and a flip set $A$ is defined as
	\begin{equation}
		x_{i}(t+1)=f_{i}\bigg(\eta_{A(t)}\Big(x(t)\Big), u(t)\bigg),\ i = 1,...,n.
		\label{eqFlipBCN}
	\end{equation}
	
	\vspace{-5pt}
	\subsection{Problems of Interests}    
	\subsubsection{Problem 1. Flipping Kernel for Reachability}
	To better illustrate Problem 1, we first define reachability.
	
	\noindent$\mathbf{Definition\ 1}$\textcolor{black}{\cite{zhou2019set}}. For system (\ref{eqFlipBCN}), let $\mathcal{M}_{0}\subset \mathcal{B}^{n}$ be the initial subset, and $\mathcal{M}_{d} \subset \mathcal{B}^{n}$ be the target subset. $\mathcal{M}_{d}$ is said to be reachable from $\mathcal{M}_{0}$ if and only if, for any initial state $x_0 \in \mathcal{M}_{0}$, there exists a sequence of joint control pairs $\Big\{\big(u(t), \eta_{A(t)}\big),t=0,1,...,T\Big\}$, such that $x_0$ reaches a target state $x_d \in \mathcal{M}_{d}$.
	
	Based on Definition 1, we define the specific problem to be considered. In some cases, it is not necessary to flip all nodes in the set $A$ for system (\ref{eqFlipBCN}) to achieve reachability. To reduce the control cost, we hope the nodes that need to be flipped as few as possible. The above problem can be transformed into finding the flipping kernel $B^*$ for reachability, which satisfies
	\begin{equation}
		\begin{aligned}
			&\min_{B}|B|\\
			\text{s.t.\ }&B\subset A \text{\textcolor{black}{\ and\ system\ (\ref{eqFlipBCN})\ from\ any\ state\ in\ }}\\
			&\mathcal{M}_{0}\text{\textcolor{black}{\ is\ reachable\ to\ state\ set\ }} \mathcal{M}_{d}.
		\end{aligned}
		\label{eqproblem1}
	\end{equation}
	It is worth noting that flipping kernel may not be unique, as there can be multiple ways to achieve reachability with the minimum cardinality $|B^*|$ through flipping.
	\subsubsection{Problem 2. Minimum Flipping Actions for Reachability}
	Based on the flipping kernel $B^*$ obtained from Problem 1 (\ref{eqproblem1}), we aim to determine the optimal policy, under which the reachability can be achieved with minimum flipping actions. This problem can be formulated as finding the policy $\pi^*:x(t)\rightarrow \big(u(t), \eta_{B^*(t)}\big),\ \forall t\in\mathbb{Z^+}$ that satisfies
	\begin{equation}
		\begin{aligned}
			\min_{\pi}\sum_{t=0}^{T}&n_t\ \ 
			\text{s.t.}\text{\textcolor{black}{\ system\ (\ref{eqFlipBCN})\ from\ any\ state\ in\ }}\\
			&\mathcal{M}_{0}\text{\textcolor{black}{\ is\ reachable\ to\ state\ set\ }} \mathcal{M}_{d},
		\end{aligned}
		\label{eqflipaction}
	\end{equation}
	where $n_t$ denotes the number of nodes to be flipped at time step $t$, and $T$ represents the terminal time step when a terminal state emerges. The terminal and initial states are represented by $\forall x_d\in\mathcal{M}_{d}$ and $\forall x_0\in\mathcal{M}_{0}$, respectively. \textcolor{black}{Note that the optimal policy $\pi^*$ may not be unique, as multiple policies can achieve reachability with minimal flipping actions.}
	
	\vspace{-5pt}
	\section{Preliminaries}
	In this section, we introduce a reinforcement learning method\textcolor{black}{, specifically focusing on the Markov decision process and the $Q$-learning algorithm.}

	\vspace{-5pt}
	\subsection{Markov Decision Process}
	Markov decision process provides the framework for reinforcement learning, which is represented as a quintuple $(\mathbf{X}, \mathbf{A}, \gamma, \mathbf{P}, \mathbf{R})$.  $\mathbf{X}=\{x_t,t\in\mathbb{Z^+}\} $ and $\mathbf{A}=\{a_t,t\in\mathbb{Z^+}\} $ denote the state space and action space, respectively. The discount factor $\gamma\in[0,1] $ weighs the importance of future rewards. The state-transition probability $\mathbf{P}_{x_t}^{x_{t+1}}(a_t)=\Pr\left\{x_{t+1} \mid x_t, a_t\right\}$ represents the chance of transitioning from state $x_t$ to $x_{t+1}$ under action $a_t$. $\mathbf{R}_{x_t}^{x_{t+1}}(a_t)=\mathbb{E}\left[r_{t+1} \mid x_t, a_t,x_{t+1}\right]$ denotes the expected reward obtained by taking action $a_t$ at state $x_t$ that transitions to $x_{t+1}$. At each time step $ t\in \mathbb{Z^+}$, an agent interacts with the environment to determine an optimal policy. Specificly, the agent observes $x_{t}$ and selects $a_t$, according to the policy $\pi:x_t\rightarrow a_t, \forall t\in\mathbb{Z^+}$. Then, the environment returns $r_{t+1}$ and $x_{t+1}$. The agent evaluates the reward $r_{t+1}$ received for taking action $a_t$ at state $x_t$ and then updates its policy $\pi$. Define $G_t=\sum_{i=t+1}^{T} \gamma^{i-t-1} r_t$ as the return. The goal of the agent is to learn the optimal policy $\pi^*:x_t\rightarrow a_t, \forall t\in\mathbb{Z^+}$, which satisfies
	\begin{equation}
		\pi^* = \max_{\pi\in\Pi}\mathbb{E}_\pi[G_t], \forall t\in\mathbb{Z^+},
	\end{equation}
	where $\mathbb{E}_\pi$ is the the expected value operator following policy $\pi$, and $\Pi$ is the set of all admissible policies. The state-value and the action-value are $v_\pi(x_t) = \mathbb{E}_\pi[G_t|x_t]$ and $q_\pi(x_t,a_t) =\mathbb{E}_\pi[G_t|x_t,a_t]$, respectively.
	The Bellman equations reveal the recursion of $v_\pi(x_t)$ and $q_\pi(x_t,a_t)$, which are given as follows:
	\begin{equation}
		\begin{aligned}
			&v_\pi(x_t) = \sum\limits_{x_{t+1}\in\mathbf{X}}\mathbf{P}_{x_t}^{x_{t+1}}\big(\pi(x_{t})\big)[\mathbf{R}_{x_t}^{x_{t+1}}\big(\pi(x_{t})\big)+\gamma v_\pi(x_{t+1})],\\
			&q_\pi(x_t,a_t)=\sum\limits_{x_{t+1}\in\mathbf{X}}\mathbf{P}_{x_t}^{x_{t+1}}(a_t)[\mathbf{R}_{x_t}^{x_{t+1}}(a_t)+\gamma v_\pi(x_{t+1})].
		\end{aligned}
		\label{eqBL}
	\end{equation} 
	
	The optimal state-value and action-value are defined as
	\begin{equation}
		\begin{aligned}
			&v^*(x_t) = \max\limits_{\pi\in\Pi} v_\pi(x_t),\forall x_t\in \mathbf{X},\\
			&q^*(x_t,a_t) = \max\limits_{\pi\in\Pi} q_\pi(x_t,a_t),\forall x_t\in \mathbf{X},\forall a_t\in \mathbf{A}.
		\end{aligned}
	\end{equation}
	
	For problems under the framework of Markov decision process, the optimal policy $\pi^* $ can be obtained through $Q$L. In the following, we introduce $Q$L. 
	
	\subsection{$Q$-Learning}
	$Q$L is a classical algorithm in reinforcement learning. The purpose of $Q$L is to enable the agent to find an optimal policy $\pi^*$ through its interactions with the environment. $Q_{t}(x_t, a_t)$, namely, the estimate of $q^*(x_t, a_t)$, is recorded and updated every time step as follows:
	\begin{equation}
		Q_{t+1}(x, a)=\left\{
		\begin{aligned}
			&Q_{t}(x, a)+\alpha_t TDE_{t},&&\text{if\ }(x, a)=\left(x_{t}, a_{t}\right),\\
			&Q_{t}(x, a), &&\text{else},
		\end{aligned}\right.
		\label{eqQupdate}
	\end{equation}
	where $\alpha_t \in (0,1], t\in \mathbb{Z^+}$ is the learning rate, and $TDE_{t}= r_{t+1}+\gamma\max\limits_{a\in\mathbf{A}}Q_t(x_{t+1},a)-Q_t(x_t,a_t)$ .
	
	In terms of action selection, $\epsilon $-greedy method is used
	\begin{equation}
		a_{t}=\left\{\begin{aligned}
			&\arg \max\limits_{a \in \mathbf{A}} Q_{t}\left(x_{t}, a\right), &&P=1-\epsilon,\\
			&\operatorname{rand}(\mathbf{A}) ,&&P=\epsilon,\\
		\end{aligned}\right. 
		\label{eqgreedy}
	\end{equation}
	where $P$ is the probability of selecting the corresponding action, $0\leq\epsilon\leq1$, and $\operatorname{rand}(\mathbf{A}) $ is an action randomly selected from $\mathbf{A}$.
	
	\noindent$\mathbf{Lemma\ 1}$ \cite{QLcovergence}. $Q_t(x_t,a_t)$ converges to the fixed point $q^*(x_t,a_t)$ with probability one under the following conditions:
	\begin{enumerate}[itemsep=1pt,topsep=1pt,parsep=1pt]
		\item $\sum_{t=0}^{\infty} \alpha_{t}=\infty$ and $\sum_{t=0}^{\infty} \alpha_{t}^{2}<\infty $.
		\item $\operatorname{var}\left[r_{t}\right] $ is finite.
		\item $|\mathbf{X}|$ and $|\mathbf{A}|$ are finite.
		\item If $\gamma=1$, all policies lead to a cost-free terminal state.
	\end{enumerate}
	
	Once $Q$L converges, the optimal policy is obtained:
	\begin{equation}
		\pi^{*}\left(x_{t}\right)=\arg \underset{a \in \mathbf{A}}{\max } Q_t\left(x_{t}, a\right) ,\forall x_{t} \in \mathbf{X}.
	\end{equation}
	
	\section{Flipping Kernel for Reachability}
	To solve Problem 1 (\ref{eqproblem1}), we propose three algorithms: iterative $Q$L, fast iterative $Q$L, and small memory iterative $Q$L. Fast iterative $Q$L enhances convergence efficiency over iterative $Q$L. Meanwhile, small memory iterative $Q$L is designed specifically for large-scale BCNs (\ref{eqFlipBCN}).

	\subsection{Markov Decision Process for Finding Flipping Kernel}
	The premise for using $Q$L is to structure the problem within the framework of Markov decision process. We represent Markov decision process by the quintuple $(\mathcal{B}^n, \mathcal{B}^{ m +|B|},\mathbf{P}, \mathbf{R},\gamma) $, where  $\mathcal{B}^n $ and $\mathcal{B}^{ m +|B|}$ are the state space and action space, respectively. The state and the action are defined as $x_t=x(t) $ and $a_t=\big(u(t), \eta_{A(t)}\big)$, respectively. The reward is given as follows:
	\begin{equation}
		r_{t}(x_t,a_t)=\left\{\begin{aligned}
			&100, &&x_{t}\in\mathcal{M}_{d},\\
			&0, &&\text{else}.\\
		\end{aligned}\right. 
		\label{eqreward}
	\end{equation}
	To incentivize the agent to reach $x_{t}\in\mathcal{M}_{d}$ as quickly as possible, we set the discount factor $\gamma\in(0,1)$.
	
	\textcolor{black}{Assuming the agent is aware of the dimensions of $\mathcal{B}^n$ and $\mathcal{B}^{m+|B|}$ but lacks knowledge of $\mathbf{P}$ and $\mathbf{R}$, meaning the agent does not comprehend the system dynamics (\ref{eqFlipBCN}). Through interactions with the environment, the agent implicitly learns about $\mathbf{P}$, $\mathbf{R}$, refines estimation of $q^{*}\left(x_{t}, a_{t}\right)$, and enhances its understanding of $\pi^{*}$.}
	
	\vspace{-5pt}
	\subsection{Iterative Q-Learning for Finding Flipping Kernel}
	Under the reward setting (\ref{eqreward}), it is worthwhile to investigate methods for determining the reachability of BCNs defined by equation (\ref{eqFlipBCN}). To answer this question, we present Theorem 1.
	
	\noindent$\mathbf{Theorem\ 1}$. For system (\ref{eqFlipBCN}), assume that $Q_0(x_t,a_t)= 0,\forall x_t,\forall a_t$ and equation (\ref{eqQupdate}) is utilized for updating $Q_t(x_t,a_t)$. \textcolor{black}{System (\ref{eqFlipBCN}) from any state in $\mathcal{M}_0$ is reachable to state set $\mathcal{M}_d$} if and only if, there exists a $t^*$ such that $\max\limits_{a\in\mathbf{A}} Q_{t^*}(x_{0}, a)>0, \forall x_0\in \mathcal{M}_{0}$.
	
	\noindent$\mathbf{Proof}$. (Necessity) Assume that \textcolor{black}{system (\ref{eqFlipBCN}) from any state in $\mathcal{M}_0$ is reachable to state set $\mathcal{M}_d$}. Without loss of generality, suppose that there is only one $x_0\in\mathcal{M}_{0}$ and one $x_d\in\mathcal{M}_{d}$. Then, there exists a joint control pair sequence $\Big\{a_t=\big(u(t), \eta_{A(t)}\big),t=0,1,...,T\Big\}$, such that $x_0$ will reach the target state $x_d$. We apply $\Big\{a_t=\big(u(t), \eta_{A(t)}\big),t=0,1,...,T\Big\}$ to $x_0$, then we obtain a trajectory $x_0\stackrel{a_0}{\longrightarrow}x_1\stackrel{a_1}{\longrightarrow}...x_T\stackrel{a_T}{\longrightarrow}x_{T+1}$, where $x_{T+1}=x_{d}$. Meanwhile, $x_t\stackrel{a_t}{\longrightarrow}x_{t+1}$ means that $x_t$ will transform into $x_{t+1}$ when $a_t$ is taken. 
	
	As $Q$L iterates, all state-action-state pairs $(x_t,a_t,x_{t+1})$ are constantly visited \cite{QLcovergence}. Therefore, there can be a case where the state-action-state pairs $(x_T,a_T,x_{T+1})$,  $(x_{T-1}, a_{T-1},x_{T})$, $(x_{T-2}, a_{T-2},x_{T-1})$, ...,$(x_0,a_0,x_{1})$ are visited one after another, and in between, other state-action-state pairs can also exist. This case causes the values in the $Q$-table to transition from all zeros to some being positive. Furthermore, we assume that this case initially occurs at time step $t'$. Then, the corresponding change of action-value from zero to positive is given by  $Q_{t'}\big(x_{T},a_T\big)\leftarrow\alpha_{t'-1}(r_{t'}+\gamma\max\limits_{a\in A}Q_{t'-1}(x_{T+1}, a))+(1-\alpha_{t'-1})Q_{t'-1}\big(x_{T},a_T\big)$, where $r_{t'}=100$ since $x_{T+1}\in\mathcal{M}_{d}$. Additionally, due to the absence of negative rewards and the fact that initial action-values are all 0, it follows that $Q_{t'-1}(x_{T+1}, a)\geq 0$ for all $a\in\mathbf{A}$. Thus, after the update, we have $Q_{t'}\big(x_{T},a_T\big)>0$. According to (\ref{eqQupdate}), at $t'+1$, it follows that $Q_{t'+1}\big(x_{T},a_T\big)>0$. Similarly, for any $ t''>t'$, it follows that $Q_{t''}\big(x_{T},a_T\big)>0$.
	
	Next, we prove that there exists $t^*$ such that $Q_{t^*}\big(x_{\overline{t}},a_{\overline{t}}\big)>0$, where $\overline{t}=T,\ T-1,\ ...,\ 0$, based on mathematical induction. The proof is divided into two parts. Firstly, there exists $t'$ such that $Q_{t'}\big(x_{\overline{t}},a_{\overline{t}}\big)>0$ where $\overline{t}=T$, according to the proof in the previous paragraph. Second, we prove that if $(x_{\overline{t}},a_{\overline{t}},x_{{\overline{t}}+1})$ is visited at time step $t''$ and $Q_{t''}\big(x_{{\overline{t}}+1},a_{{\overline{t}}+1}\big)>0$, then $Q_{t''+1}\big(x_{\overline{t}},a_{\overline{t}}\big)>0$. Since $(x_{\overline{t}},a_{\overline{t}},x_{{\overline{t}}+1})$ is visited at time step $t''$, it follows that $Q_{t''+1}\big(x_{\overline{t}},a_{\overline{t}}\big)\leftarrow\alpha_{t''}(r_{t''+1}+\gamma\max\limits_{a\in\mathbf{A} }Q_{t''}(x_{{\overline{t}}+1},a))+(1-\alpha_{t''})Q_{t''}\big(x_{{\overline{t}}},a_{{\overline{t}}}\big)$. In the updated formula, we have $r_{t''+1}\geq0$, and $Q_{t''}\big(x_{{\overline{t}}},a_{{\overline{t}}}\big)\geq0$. Furthermore, since $Q_{t''}(x_{{\overline{t}}+1},a_{{\overline{t}}+1})>0$,  it implies that $\max\limits_{a\in\mathbf{A}}Q_{t''}(x_{{\overline{t}}+1},a)>0$. Consequently, we have $Q_{t''+1}\big(x_{\overline{t}},a_{\overline{t}}\big)>0$. According to mathematical induction, there exists $t^*$ such that $Q_{t^*}\big(x_{0},a_0\big)>0$. Since $a_0\in\mathbf{A}$ holds, we have $\max\limits_{a\in\mathbf{A}} Q_{t^*}(x_{0}, a)>0$. 
	
	(Sufficiency) Suppose that there exists $t^*$ such that $\max\limits_{a\in\mathbf{A}} Q_{t^*}(x_{0}, a)>0,\ \forall x_0\in\mathcal{M}_{0}$. Without loss of generality, we assume that there is only one $x_0\in\mathcal{M}_{0}$ as well as one $x_d\in\mathcal{M}_{d}$, $\text{arg}\max\limits_{a\in\mathbf{A}} Q_{t^*}(x_{0}, a)=a_t $, and $\max\limits_{a\in\mathbf{A}} Q_{t^*-1}(x_{0}, a)=0$. Then, according to the updated formula $Q_{t^*}\big(x_{0},a_{t}\big)\leftarrow\alpha_{t^*-1}(r_{t^*}+\gamma\max\limits_{a\in\mathbf{A}}Q_{t^*-1}(x_{0\text{next}},a))+(1-\alpha_{t^*-1})Q_{t^*-1}\big(x_{0},a_{t}\big)$, where $x_{0\text{next}}$ is the subsequent state of $x_0$, at least one of the following two cases exists:
	
	\begin{itemize}[itemsep=0.85pt,topsep=0.85pt,parsep=0.85pt]
		
		\item Case 1. There exist $a_t$ and $x_{0\text{next}}$ such that $r_{t^*}(x_{0\text{next}},a_t)$ $>0$.
		\item Case 2. $x_{0\text{next}}$ satisfies $\max\limits_{a\in\mathbf{A}} Q_{t^*-1}(x_{0\text{next}}, a)>0$. 
	\end{itemize}
	
	In Case 1, we have $x_{0\text{next}}\in\mathcal{M}_{d}$.
	\textcolor{black}{It implies that system (\ref{eqFlipBCN}) from state $x_0$ is reachable to state set $\mathcal{M}_d$} in one step. 
	
	In terms of Case 2, let us consider the condition under which $\max\limits_{a\in\mathbf{A}} Q_{t^*-1}(x_{0\text{next}}, a)>0$, meaning that there exists an action $a\in\mathbf{A}$ such that  $Q_{t^*-1}(x_{0\text{next}}, a)>0$. We define the $i^{th}$ action-value, which changes from 0 to a positive number, as $Q_{t_i}(x_i,a_i)$. Here, $t_i$ denotes the time step of the change of the $i^{th}$ action-value, and $(x_i,a_i)$ corresponds to the state-action pair. Since the initial $Q$-values are all 0, it follows that $Q_{t_1}(x_1,a_1)>0$ if and only if $x_{1\text{next}}\in\mathcal{M}_{d}$. This implies that \textcolor{black}{system (3) from state $x_1$ is reachable to state set $\mathcal{M}_d$} 
	within one step, leading to $r_{t_1}>0$. Next, for any $i>1$, $Q_{t_i}(x_i,a_i)>0$ occurs if and only if $x_{i\text{next}}\in\mathcal{M}_{d}$, or $x_{i\text{next}}=x_j,\ j<i$, where $x_j$ is a state that has satisfied $Q_{t_j}(x_j,a_j)>0$ for $t_j<t_i$ based on the previously defined conditions. This means that for $Q_{t_i}(x_i,a_i)>0,\ \forall i$, at least one of the following two events has taken place: either $\mathcal{M}_{d}$ is reachable from $x_i$; or a state $x_j,j<i$, which can reach $\mathcal{M}_{d}$, is reachable from $x_i$. In both cases, it implies that $\mathcal{M}_{d}$ is reachable from $x_i$. Thus, if $Q_{t^*-1}(x_{0\text{next}}, a)>0$, it follows that $\mathcal{M}_{d}$ is reachable from $x_{0\text{next}}$. This, in turn, implies that $\mathcal{M}_{d}$ is also reachable from $x_{0}$. Based on the analysis of Cases 1 and 2, if $\max\limits_{a\in\mathbf{A}} Q_{t^*-1}(x_{0}, a)$ $>0$ holds for all $ x_0\in \mathcal{M}_{0}$, then $\mathcal{M}_{d}$ is reachable from $\mathcal{M}_{0}$.  $\hfill\blacksquare$
	
	\noindent $\mathbf{Remark\ 1}$. The key for $Q$L to judge the reachability for system (\ref{eqFlipBCN}) lies in exploration (\ref{eqgreedy}) and update rule (\ref{eqQupdate}).
	
	\noindent $\mathbf{Lemma\ 2}$. For system (\ref{eqFlipBCN}),  $\mathcal{M}_{d}$ is reachable from $\mathcal{M}_{0}$ if and only if there exists $l\leqslant 2^{n}-\left|\mathcal{M}_{d}\right|$ such that $\mathcal{M}_{d}$ is $l$-step reachable from $\forall x_0\in \mathcal{M}_{0}$.
	
	\noindent $\mathbf{Proof}$. Without loss of generality, let us assume that there is a trajectory from $x_0\in\mathcal{M}_{0}$ to $\mathcal{M}_{d}$ whose length is greater than $2^{n}-\left|\mathcal{M}_{d}\right|$. Then, there must be at least one $x_t\in\mathcal{B}^n\setminus\mathcal{M}_{d}$ that is visited repeatedly. If we remove the cycle from $x_t$ to $x_t$, the trajectory length will be less than $2^{n}-\left|\mathcal{M}_{d}\right|$. $\hfill\blacksquare$
	
	Based on Theorem 1 and Lemma 2, we present Algorithm 1 for finding flipping kernels. \textcolor{black}{First, we provide the notation definitions used in Algorithm 1. $C_{|A|}^k$ denotes the combinatorial number of flip sets with the given $k$ nodes in $A$. $B_{k_i}$ represents the $i^{th}$ flip set with the given $k$ nodes. $ep$ denotes the number of episodes, where ``Episode" is a term from reinforcement learning, referring to a period of interaction that has passed through $T_{max}$ time steps from any initial state $x_0\in \mathcal{M}_{0}$. $T_{max}$ is the maximum time step, which should exceed the length of the trajectory from $\mathcal{M}_{0}$ to $\mathcal{M}_{d}$. One can refer to Lemma 2 for the setting of $T_{max}$. }
	
	\textcolor{black}{The core idea of Algorithm 1 is to incrementally increase the flip set size from $|B|$ to $|B|+1$ in step 18 when reachability isn't achieved with flip sets of size $|B|$. Specifically, step 4 fixes the flip set $B$, followed by steps 5-16 assessing reachability under that flip set, guided by Theorem 1. Upon identifying the first flipping kernel (when the condition in step 12 is met), the cardinality of the kernel is determined. This leads to setting $K=|B|$ as per step 13, along with step 2 to prevent ineffective access to a larger flip set. After evaluating system reachability under all flip sets with $K$ nodes, steps 20-24 present the algorithm's results.}
	
	
	\begin{algorithm}[htb] 
		\caption*{ $\mathbf{Algorithm\ 1}$ Finding flipping kernels for reachability of BCNs under state-flipped control using iterative $Q$L} 
		\label{algKQL} 
		\begin{algorithmic}[1] 
			\REQUIRE  
			$\mathcal{M}_{0}$, $\mathcal{M}_{d}$, $A$, $\alpha_{t}\in (0,1]$, $\gamma\in (0,1)$, $\epsilon\in[0,1]$, $N$, $T_{max}$\\
			\ENSURE  
			Flipping kernels
			\STATE $K = |A|$, $k=0$, $n = 0$
			\WHILE{ $k \leq K$ }
			\FOR{ $i = 1,2,\dots,C_{|A|}^k$ }
			\STATE $B=B_{k_i}$
			\STATE Initialize $Q(x_t,a_t) \leftarrow 0,\forall x_t\in \mathcal{B}^n ,\forall a_t\in \mathcal{B}^{ m +|B|}$
			\FOR{ $ep = 0, 1,\dots, N-1 $ }
			\STATE $x_0 \leftarrow \operatorname{rand}(\mathcal{M}_{0})$
			\FOR{$t=0,1,\dots, T_{max}-1 $ and $x_t\notin\mathcal{M}_{d}$}
			\STATE $a_{t}\leftarrow\left\{
			\begin{aligned}
				&\arg \max _{a_t\in \mathcal{B}^{ m +|B|}} Q(x_{t}, a_t),&&P=1-\epsilon\\
				&\operatorname{rand}(\mathcal{B}^{ m +|B|}),&&P=\epsilon
			\end{aligned}\right. $
			\STATE$Q(x_t,a_t)\leftarrow\alpha_{t}(\gamma\max\limits_{a_{t+1}\in \mathcal{B}^{ m +|B|}}Q(x_{t+1}, a_{t+1})$\ $+r_{t+1})+(1-\alpha_{t})Q(x_t,a_t)$
			\ENDFOR 
			\IF{$\max\limits_{a_t\in \mathcal{B}^{ m +|B|}} Q(x_{t}, a_t)>0, \forall x_t\in \mathcal{M}_{0}$}
			\STATE $B_n^*=B$, $K = |B|$, $n=n+1$
			\STATE Break
			\ENDIF
			\ENDFOR
			\ENDFOR  
			\STATE $k = k+1$
			\ENDWHILE
			\IF{$n=0$}
			\RETURN ``System can't achieve reachability."
			\ELSE
			\RETURN $B^*_1,\dots,B^*_n$
			\ENDIF
		\end{algorithmic}
	\end{algorithm}
	
	\textcolor{black}{Although Algorithm 1 is efficient, there is room for improvement. The next section outlines techniques to enhance its convergence efficiency.}

	
	\vspace{-5pt}
	\subsection{Fast Iterative Q-learning for Finding Flipping Kernel}
	In this subsection, we propose Algorithm 2, known as fast iterative $Q$L for finding the flipping kernels, which improves the convergence efficiency. The main difference between Algorithm 2 and Algorithm 1 can be classified into two aspects. 1) Special initial states: Algorithm 2 selects initial states strategically instead of randomly. 2) TL: Algorithm 2 utilizes the knowledge gained from achieving reachability with smaller flip sets to search for flipping kernels for larger flip sets.
	
	\begin{algorithm}[htb] 
		\caption*{ $\mathbf{Algorithm\ 2}$ Finding flipping kernels for reachability of  BCNs under state-flipped control using fast iterative $Q$L} 
		\label{algKQL} 
		\begin{algorithmic}[1] 
			\REQUIRE  
			$\mathcal{M}_{0}$, $\mathcal{M}_{d}$, $A$, $\alpha_{t}\in (0,1]$, $\gamma\in (0,1)$, $\epsilon\in[0,1]$, $N$, $T_{max}$\\
			\ENSURE  
			Flipping kernels
			\STATE $K = |A|$, $k=0$, $n = 0$
			\WHILE{ $k \leq K$ }
			\FOR{ $i = 1,2,\dots,C_{|A|}^k$ }
			\STATE $B=B_{k_i}$
			\STATE Initialize $Q(x_t,a_t),\forall x_t\in \mathcal{B}^n ,\forall a_t\in \mathcal{B}^{ m +|B|}$ using \textcolor{black}{equation} (\ref{eqtrans})
			\STATE Examining the reachability under flip set $B$ following \textcolor{black}{\textbf{modified steps 6-16 of Algorithm 1} with equation (\ref{eqstep7})}
			\STATE Record the $Q$-table for $B$
			\ENDFOR  
			\STATE Drop all the $Q$-tables, $k = k+1$
			\ENDWHILE
			\STATE \textcolor{black}{Present the result following \textbf{steps 20-24 of Algorithm 1}}
		\end{algorithmic}
	\end{algorithm}
	
	\subsubsection{Special Initial States}
	The main concept behind selecting special initial states is to avoid visiting $x_0\in\mathcal{M}_{0}$ once we determine that $\mathcal{M}_{d}$ is reachable from it. Instead, we focus on states in $\mathcal{M}_{0}$ that have not been identified as reachable. \textcolor{black}{With this approach in mind, we modify step 7 in Algorithm 1 into 
		\begin{equation}
			x_{0} \leftarrow \operatorname{rand}\left(\left\{x_{0} \in \mathcal{M}_{0} \mid \max _{a_{t} \in \mathcal{B}^{m+|B|}} Q\left(x_{0}, a_{t}\right)=0\right\}\right)
			\label{eqstep7}
	\end{equation}}
	
	\noindent $\mathbf{Remark\ 2.}$ The validity of Theorem 1 remains when special initial states are added to iterative $Q$L, as the trajectory from the states in $\mathcal{M}_{0}$ that are reachable to $\mathcal{M}_{d}$ can still be visited.
	
	\subsubsection{Transfer-learning}
	Let $Q^B(x_t,a_t)$ and $Q^b(x_t,a_t)$ represent the $Q$-table with flip sets $B$ and $b$ respectively, where $b\subset B$. The relationship between $Q^B(x_t,a_t)$ and $Q^b(x_t,a_t)$ is explained in Theorem 2 below.
	
	\noindent$\mathbf{Theorem\ 2}$. For any state-action pairs $(x_t,a_t)$, if $Q^b(x_t,a_t)$ $>0$ with flip set $b\subset B$, then $Q^B(x_t,a_t)>0$ holds for flip set $B$.
	
	\noindent$\mathbf{Proof}$. It is evident that the action space for flip set $b\subset B$ is the subset of the action space for flip set $B$, namely, $ \mathbf{A}_b\subset\mathbf{A}_B$. Therefore, if 
	\textcolor{black}{system (\ref{eqFlipBCN}) from state $x_1$ is reachable to state $x_d\in\mathcal{M}_{d}$}  with flip set $b\subset B$, namely, there exists a \textcolor{black}{trajectory} $x_1\stackrel{a_1}{\longrightarrow}x_2...\stackrel{a_T}{\longrightarrow}x_d,\ a_t\in\mathbf{A}_b$, the reachability still holds with flip set $B$. This is because the \textcolor{black}{trajectory} $x_1\stackrel{a_1}{\longrightarrow}x_2...\stackrel{a_T}{\longrightarrow}x_d$  exists for $ a_t\in\mathbf{A}_b\subset\mathbf{A}_B$. Thus, if $\mathcal{M}_{d}$  is reachable from $x_1$ with flip set $b$, then the reachability holds with flip set $B$. According to Theorem 1, for all $(x_t,a_t)$, if $Q^b(x_t,a_t)>0$ with flip set $b$, then $Q^B(x_t,a_t)>0$ holds for flip set $B$.
	$\hfill\blacksquare$
	
	Motivated by the above results, we consider TL \cite{transfer}. TL enhances the learning process in a new task by transferring knowledge from a related task that has already been learned. In this context, we employ TL by initializing the $Q$-table for flip set $B$ with the knowledge gained from the $Q$-tables for flip sets $b\subset B$, namely
	\begin{equation}
		Q_{0}(x_t, a_t)=\left\{
		\begin{aligned}
			&0, \text{\ if\ no\ subset\ } b\subset B\text{\ such\  that\ } a_t\in b,\\
			&\max_{b\subset B}Q_{t}^b(x_t,a_t), \text{\ else},
		\end{aligned}\right.
		\label{eqtrans}
	\end{equation}
	where $Q_{t}^b(x_t,a_t)$ represents the $Q$-table associated with flip set $b$. \textcolor{black}{Based on this idea, we refine step 5 and include steps 7 and 9 in Algorithm 2 compared to Algorithm 1.}
	
	Algorithm 2 improves the convergence efficiency compared to Algorithm 1. However, Algorithm 2 is not suitable for large-scale BCNs. Specifically, the operation of iterative $Q$L is based on a $Q$-table that has $|\mathbf{X}|\times|\mathbf{A}| $ values. The number of values in the table grows exponentially with $n$ and $m+|B|$, and can be expressed as $2^{n+m+|B|}$. When $n$ and $m+|B|$ are large, iterative $Q$L becomes impractical because the $Q$-table is too large to store on a computer. Therefore, we propose an algorithm in the following subsection, which can be applied to large-scale BCNs.
	
	\vspace{-5pt}
	\subsection{Small Memory Iterative Q-Learning for Finding Flipping Kernel of Large-scale BCNs}
	In this subsection, we present Algorithm 3, named small memory iterative $Q$L, which is inspired by the work of \cite{pxl}. This algorithm serves as a solution for identifying flipping kernels in large-scale BCNs. The core idea behind Algorithm 3 is to store action-values only for visited states, instead of all $x_t\in\mathcal{B}^{n} $, to reduce memory consumption. To implement this idea, we make adjustments to step 5 in Algorithm 1 and introduce additional steps 10-12 in Algorithm 3. Subsequently, we offer detailed insights into the effectiveness of Algorithm 3 and its applicability.
	
	\begin{algorithm}[htb] 
		\caption*{ $\mathbf{Algorithm\ 3}$ Finding flipping kernels for reachability of \textcolor{black}{system (\ref{eqFlipBCN})} using small memory iterative $Q$L} 
		\label{algKQL} 
		\begin{algorithmic}[1] 
			\REQUIRE 
			$\mathcal{M}_{0}$, $\mathcal{M}_{d}$, $A$, $\alpha_{t}\in (0,1]$, $\gamma\in (0,1)$, $\epsilon\in[0,1]$, $N$, $T_{max}$\\
			\ENSURE  
			Flipping kernels
			\STATE $K = |A|$, $k=0$, $n = 0$
			\WHILE{ $k \leq K$ }
			\FOR{ $i = 1,2,\dots,C_{|A|}^k$ }
			\STATE $B=B_{k_i}$
			\STATE Initialize $Q(x_t,a_t) \leftarrow 0,\forall x_t\in \mathcal{M}_{0} ,\forall a_t\in \mathcal{B}^{ m +|B|}$
			\FOR{ $ep = 0, 1,\dots, N-1 $ }
			\STATE $x_0 \leftarrow \operatorname{rand}(\mathcal{M}_{0})$
			\FOR{$t=0,1,\dots, T_{max}-1 $ and $x_t\notin\mathcal{M}_{d}$}
			\STATE $a_{t}\leftarrow\left\{
			\begin{aligned}
				&\arg \max\limits_{a_t\in \mathcal{B}^{ m +|B|}} Q(x_{t}, a_{t}),&&P=1-\epsilon\\
				&\operatorname{rand}(\mathcal{B}^{ m +|B|}),&&P=\epsilon
			\end{aligned}\right. $
			\IF {$Q(x_{t+1},.)$ is not in the $Q$-table}
			\STATE Add $Q(x_{t+1},.)=0$ to the $Q$-table
			\ENDIF
			\STATE$Q(x_t,a_t)\leftarrow\alpha_{t}(\gamma\max\limits_{a_{t+1}\in \mathcal{B}^{ m +|B|}}Q(x_{t+1}, a_{t+1})$\ $+r_{t+1})+(1-\alpha_{t})Q(x_t,a_t)$
			\ENDFOR 
			\IF{$\max\limits_{a_t\in \mathcal{B}^{ m +|B|}} Q(x_{t}, a_t)>0, \forall x_t\in \mathcal{M}_{0}$}
			\STATE $B_n^*=B$, $K = |B|$, $n=n+1$
			\STATE Break
			\ENDIF
			\ENDFOR
			\ENDFOR  
			\STATE $k = k+1$
			\ENDWHILE
			\STATE \textcolor{black}{Present the result following \textbf{steps 20-24 of Algorithm 1}}
		\end{algorithmic}
	\end{algorithm}

	Since Algorithm 3 only stores action-values for visited states, it reduces the number of required action-values. Specifically, let's define $V=\{x|$ $x$ is reachable from $\mathcal{M}_{0}$, $ x\in\mathcal{B}^n\}$. Then, we can observe that \textcolor{black}{$|V\cup\mathcal{M}_{0}|\times2^{m+|B|}$} is the total number of visitable states, \textcolor{black}{where $\cup$ represents the union of sets.} Therefore, we can conclude that the number of required action-values using Algorithm 3 is \textcolor{black}{$|V\cup\mathcal{M}_{0}|\times2^{m+|B|}$}, which is a number not exceeding $2^{n+m+|B|}$. Next, we present Theorem 3 to provide an estimation of $|V|$. To better understand the theorem, we first introduce Definition 2.
	
	\noindent $\mathbf{Definition\ 2}$\textcolor{black}{\cite{flip6}}. The in-degree of $x\in\mathcal{B}^n$ is defined as the number of states which can reach $x$ in one step with $\overline{a_t}=\big(u(t), \eta_A(t)\big)$ where $|\eta_{A(t)}|_\infty=0$.
	
	\noindent $\mathbf{Theorem\ 3}$. Define $I=\{x|\ $the in-degree of $x$ is greater than 0, $x\in\mathcal{B}^n\}$. Then, it follows that $|V|\leq|I|$.
	
	\noindent $\mathbf{Proof}$. The state transition process can be divided into two steps. For any $x'\in\mathcal{B}^n$, according to $\eta_{A(t)}$, it is firstly flipped into $\eta_{A(t)}(x')\in\mathcal{B}^n$. Second, based on $u(t)$, namely, $\overline{a_t}$, $\eta_{A(t)}(x')\in\mathcal{B}^n$ tranfers to $x\in\mathcal{B}^n$. According to the second step of the state transition process and the definition of $I$, we have $x\in I$. Due to the arbitrariness of $x'$, it follows that $\{x|\ x$ is reachable from $\mathcal{B}^n$ in one step, $ x\in\mathcal{B}^n\}\subset I$. Since $\mathcal{B}^n$ contains all states, the condition `in one step' can be removed. Define $N=\{x|\ x$ is reachable from $\mathcal{B}^n$, $ x\in\mathcal{B}^n\}$. So, we have $N\subset I$. Notice that the only difference between the sets $N$ and $V$ is the domain of the initial states. Since we have $\mathcal{M}_{0}\subset\mathcal{B}^n$, it follows that $V\subset N \subset I$. Therefore, it can be concluded that $|V|\leq|I|$. $\hfill\blacksquare$
	
	Algorithm 3 can reduce the number of entries stored in the $Q$-table from $2 ^ {n+m + | B |} $ to \textcolor{black}{$| V\cup\mathcal{M}_{0} |\times2 ^ {m + | B |} $}, where $|V|\leq|I|$. However, we acknowledge that when the size of \textcolor{black}{$|V\cup\mathcal{M}_{0} |\times2 ^ {m + | B |} $} becomes excessively large, $Q$L faces certain limitations. It is important to note that Algorithm 3 offers a potential solution for addressing problems in large-scale BCNs, as opposed to the conventional $Q$L approach, which is often deemed inapplicable. In fact, the value of \textcolor{black}{$|V\cup\mathcal{M}_{0}|$} is not directly related to $n$. Specifically, large-scale BCNs with low connectivity will result in a small number of \textcolor{black}{$|V\cup\mathcal{M}_{0}|$}. Traditional $Q$L preconceives to store $2 ^ {n+m + | B |} $ values in the $Q$-table, which not only restricts the ability to handle problems in large-scale systems but also lacks practicality. In contrast, Algorithm 3 allows agents to explore and determine the significant information that needs to be recorded. This approach provides an opportunity to solve the reachability problems of large-scale BCNs.
	
	\noindent $\mathbf{Remark\ 3}$. The idea of utilizing special initial states and TL in Algorithm 2 can also be incorporated into Algorithm 3. The hybrid algorithm proposed has two major advantages: improved convergence efficiency and expanded applicability to large-scale BCNs.

	\vspace{-5pt}
	\section{Minimum Flipping Actions for Reachability}
	In this section, we aim to solve Problem 2 (\ref{eqflipaction}). First, we propose a reward setting in which the highest return is achieved only when the policy successfully reaches the goal. Then, we propose two algorithms for obtaining the optimal policies: $Q$L for small-scale BCNs, and small memory $Q$L for large-scale ones.
	
	\vspace{-5pt}
	\subsection{Markov Decision Process for Minimizing Flipping Actions}
	Since $Q$L will be used to find $\pi^*$, we first construct the problem into the framework of Markov decision process. Let Markov decision process be the quintuple $(\mathcal{B}^n, \mathcal{B}^{ m +|B^*|}, \gamma,\mathbf{P}, \mathbf{R})$. To guide the agent in achieving the goal with minimal flipping actions, we define $r_{t}$ as follows:
	\begin{equation}
		r_{t}(x_t,a_t)=\left\{\begin{aligned}
			&-w\times n_t, &&x_{t}\in\mathcal{M}_{d},\\
			&-w\times n_t-1, &&\text{else},\\
		\end{aligned}\right. 
		\label{eqr2}
	\end{equation}
	where $w>0$ represents the weight. The reward $r_{t}$ consists of two parts. The first component is ``-1'', which is assigned when the agent has not yet reached $\mathcal{M}_{d}$. This negative feedback encourages the agent to reach the goal as soon as possible. The second component is ``$-w\times n_t$'', which incentivizes the agent to use as few flipping actions as possible to reach the goal. Since we aim to minimize the cumulative flipping actions at each time step, we set $\gamma=1$ indicating that future rewards are not discounted in importance.
	
	\noindent$\mathbf{Remark\ 4}$. The following analysis is based on the assumption that \textcolor{black}{system (\ref{eqFlipBCN}) from any state in $\mathcal{M}_0$ is reachable to state set $\mathcal{M}_d$}. The validity of this assumption can be verified using Theorem 1.
	
	The selection of $w$ plays a crucial role in effectively conveying the goal (\ref{eqflipaction}) to the agent. If $w$ is too small, the objective of 
	achieving reachability as soon as possible will submerge the objective of minimizing flipping actions. To illustrate this issue more clearly, let us consider an example.
	
	\noindent$\mathbf{Example\ 1}$. Without loss of generality, we assume that there is only one $x_0\in\mathcal{M}_{0}$ and one $x_d\in\mathcal{M}_{d}$. Suppose that there exist two policies $\pi_1$ and $\pi_2$. When we start with $x_0$ and take $\pi_1$, we obtain the trajectory $x_0\rightarrow x_1\rightarrow x_2\rightarrow x_3\rightarrow x_d$ without any flipping action. If we take $\pi_2$, the trajectory $x_0\rightarrow x_d$ with 2 flipping actions will be obtained. It can be calculated that $v_{\pi_1}(x_0)=-4$, since the agent takes 4 steps to achieve reachability without flipping actions. Meanwhile, $v_{\pi_1}(x_0)=-1-2w$, since the agent takes 1 step to achieve reachability with 2 flipping actions. If we set $w=1$, then it will follow that $v_{\pi_1}(x_0)<v_{\pi_2}(x_0)$. Therefore, the obtained optimal policy $\pi_2$ requires more flipping actions to achieve reachability, which does not contribute to finding $\pi^*$ for equation (\ref{eqflipaction}).
	
	\textcolor{black}{Furthermore, although a high value of the parameter ``$w$" assists in reducing flipping actions for reachability, it can present challenges in reinforcement learning. Convergence in these algorithms relies on a finite reward variance, as specified in the second condition of Lemma 1. As ``$w$" approaches infinity, this criterion cannot be satisfied. Additionally, an excessively large value of ``$w$" may increase reward variance, resulting in greater variability in estimated state values and subsequently affecting the speed of algorithm convergence \cite{greensmith2004variance}. Hence, establishing a lower bound for ``$w$" becomes crucial to facilitate algorithm convergence while ensuring goal attainment.
	}
	
	So, what should be the appropriate value of $w$ to achieve the goal (\ref{eqflipaction})? To address this question, sufficient conditions that ensure the optimality of the policy are proposed in Theorem 4 and Corollary 1. The main concept is to select a value of $w$ such that $v_{\pi^*}(x_0)>v_{\pi}(x_0),\forall \pi \in \Pi,$ and $\pi\neq\pi^*$, where $\pi^*$ satisfies the goal (\ref{eqflipaction}).
	
	\noindent $\mathbf{Theorem\ 4}$. For system (\ref{eqFlipBCN}), if $w>l$, where $l$ is the maximum number of steps required from any initial states $x_0\in\mathcal{M}_{0}$ to reach $\mathcal{M}_{d}$ without cycles, then the optimal policy obtained based on (\ref{eqr2}) satisfies the goal (\ref{eqflipaction}). 
	
	\noindent$\mathbf{Proof}$. Without loss of generality, it is assumed that there is only one $x_0\in\mathcal{M}_{0}$. We classify all $\pi\in\Pi$ and prove that, under the condition that $w>l$, the policy with the highest state-value satisfies the goal (\ref{eqflipaction}). For all $\pi\in\Pi$, we can divide them into two cases:
	\begin{itemize}[itemsep=0.85pt,topsep=0.85pt,parsep=0.85pt]
		\item Case 1. Following $\pi_1$, \textcolor{black}{system (\ref{eqFlipBCN}) from state $x_0$ is not reachable to state set $\mathcal{M}_d$}
		
		\item Case 2. Following $\pi_2$, \textcolor{black}{system (\ref{eqFlipBCN}) from state $x_0$ is reachable to state set $\mathcal{M}_d$}
	\end{itemize}
	
	For Case 1, $v_{\pi_1}(x_0)=\sum_{t=0}^{-\infty}-1=-\infty$. The value of $v_{\pi_1}(x_0)$ is insufficient for $\pi_1$ to be optimal. For Case 2, we divide it into two sub-cases:
	
	\begin{itemize}[itemsep=0.85pt,topsep=0.85pt,parsep=0.85pt]
		\item Case 2.1. Following $\pi_{2.1}$, \textcolor{black}{system (\ref{eqFlipBCN}) from state $x_0$ is reachable to state set $\mathcal{M}_d$} with fewer flipping actions.
		\item Case 2.2. Following $\pi_{2.2}$, \textcolor{black}{system (\ref{eqFlipBCN}) from state $x_0$ is reachable to state set $\mathcal{M}_d$} with more flipping actions. 
	\end{itemize}
	
	Next, we compare the magnitude of $v_{\pi_{2.1}}$ and $v_{\pi_{2.2}}$. Let us assume that it takes $T_{\pi_{2.1}}$ steps with $\sum n_{\pi_{2.1}}$ flipping actions, and $T_{\pi_{2.2}}$ steps with $\sum n_{\pi_{2.2}}$ flipping actions for \textcolor{black}{system (\ref{eqFlipBCN}) from state} $x_0$ to reach \textcolor{black}{state set} $\mathcal{M}_{d}$ under $\pi_{2.1}$ and $\pi_{2.2}$, respectively. Then, we can obtain $v_{\pi_{2.1}}=-w\sum n_{\pi_{2.1}}-T_{\pi_{2.1}}$ and $v_{\pi_{2.2}}=-w\sum n_{\pi_{2.2}}-T_{\pi_{2.2}}$. It is calculated that 
	\begin{equation}
		v_{\pi_{2.1}}-v_{\pi_{2.2}}=w\sum n_{\pi_{2.2}}+T_{\pi_{2.2}}-w\sum n_{\pi_{2.1}}-T_{\pi_{2.1}}.
		\label{eqvminus}
	\end{equation}
	
	Since more flipping actions are taken under $\pi_{2.2}$ than $\pi_{2.1}$ to achieve reachability, then one has $\sum n_{\pi_{2.2}}-\sum n_{\pi_{2.1}}>0$. 
	
	Regarding the relationship between $T_{\pi_{2.1}}$ and $T_{\pi_{2.2}}$, there are two categories:
	
	\begin{itemize}[itemsep=0.85pt,topsep=0.85pt,parsep=0.85pt]
		\item Category 1. $T_{\pi_{2.1}}\leq T_{\pi_{2.2}}$.
		\item Category 2. $T_{\pi_{2.1}}> T_{\pi_{2.2}}$. 
	\end{itemize}
	
	For Category 1, it's obvious that $v_{\pi_{2.1}}>v_{\pi_{2.2}}$. For Category 2, we divide $\pi_{2.1}$ into two cases.
	
	\begin{itemize}[itemsep=0.85pt,topsep=0.85pt,parsep=0.85pt]
		\item Case 2.1.1. Following $\pi_{2.1.1}$, \textcolor{black}{system (\ref{eqFlipBCN}) from state} $x_0$ can reach \textcolor{black}{state set} $\mathcal{M}_{d}$ in $T_{\pi_{2.1.1}}\leq l$ steps with $\sum n_{\pi_{2.1.1}}<\sum n_{\pi_{2.2}}$ flipping actions, where $l$ is the maximum length of the trajectory from $x_0$ to $\mathcal{M}_{d}$ without cycle.  
		\item Case 2.1.2. Following $\pi_{2.1.2}$, \textcolor{black}{system (\ref{eqFlipBCN}) from state} $x_0$ can reach \textcolor{black}{state set} $\mathcal{M}_{d}$ in $T_{\pi_{2.1.2}}>l$ steps with $\sum n_{\pi_{2.1.2}}<\sum n_{\pi_{2.2}}$ flipping actions.
	\end{itemize}
	
	Next, we prove $v_{\pi_{2.1.1}}>v_{\pi_{2.2}}$. For Case 2.1.1, we have the following inequalities
	\begin{equation}
		\left\{\begin{aligned}
			&T_{\pi_{2.2}}-T_{\pi_{2.1.1}}\geq T_{min}-l\geq1-l,\\
			&\sum n_{\pi_{2.2}}-\sum n_{\pi_{2.1.1}}\geq1,\\
		\end{aligned}\right. 
		\label{eqineq}
	\end{equation}
	where $T_{min}$ is the minumum length of the trajectory from $x_0$ to $\mathcal{M}_{d}$. Substitute (\ref{eqineq}) into (\ref{eqvminus}), we have $v_{\pi_{2.1.1}}-v_{\pi_{2.2}}\geq 1-l+w$. Since $w>l$, it follows that $v_{\pi_{2.1.1}}-v_{\pi_{2.2}}>0$. 
	
	In the following, we discuss Case 2.1.2. Note that $T_{\pi_{2.1.2}}>l$ implies that at least one $x_{t} \in \mathcal{B}^{n} \backslash \mathcal{M}_{d}$ is visited repeatedly, according to Lemma 2. If we eliminate the cycle from the trajectory, then its length will be no greater than $l$, and its flipping actions are fewer. This indicates that there exists $\pi_{2.1.2}^{*}$ such that $T_{\pi_{2.1.2}^{*}}\leq l$ and $\sum n_{\pi_{2.1.2}^{*}}\leq\sum n_{\pi_{2.1.2}}<\sum n_{\pi_{2.2}}$. Thus, $\pi_{2.1.2}^{*}$ is consistent with Case 2.1.1. Based on the proof in the previous paragraph, we have $v_{\pi_{2.1.2}^{*}}(x_0)>v_{\pi_{2.2}}(x_0)$. Also, it is easy to see that $v_{\pi_{2.1.2}^{*}}(x_0)>v_{\pi_{2.1.2}}(x_0)$. Therefore, it can be concluded that $\pi_{2.1.2}^{*}$ has the highest return compared with $\pi_{2.2}$ and $\pi_{2.1.2}$. 
	
	After analyzing various cases, we deduce that when $w>l$ holds, one of the following two conditions must be satisfied: 1) $v_{\pi_{2.1}}(x_0)>v_{\pi_{2.2}}(x_0)$; 2) There exists an alternative policy $\pi^*_{2.1.2}$ that achieves reachability with fewer flipping actions compared to $\pi_{2.1}$, satisfying the conditions $v_{\pi^*_{2.1.2}}(x_0)>v_{\pi_{2.2}}(x_0)$ and $v_{\pi^*_{2.1.2}}(x_0)>v_{\pi_{2.1}}(x_0)$. Namely, $\pi$ yielding the highest $v_{\pi}(x_0)$ satisfies the goal (\ref{eqflipaction}). $\hfill\blacksquare$
	
	In Theorem 4, the knowledge of $l$ is necessary. If $l$ is unknown, we can deduce the following corollary according to Lemma 2: $l\leq2^n-|\mathcal{M}_{d}|$.
	
	\noindent$\mathbf{Corollary\ 1}$. If $w>2^n-|\mathcal{M}_{d}|$, then the optimal policy obtained according to (\ref{eqr2}) satisfies the goal (\ref{eqflipaction}). 
	
	\noindent$\mathbf{Remark\ 5}$. Two points should be clarified for Theorem 4 and Corollary 1. First, following Theorem 4 or Corollary 1, we do not obtain all $\pi^*$ that satisfy the goal (\ref{eqflipaction}), but rather the $\pi^*$ that satisfies the goal (\ref{eqflipaction}) with the shortest time to achieve reachability. Second, Theorem 4 and Corollary 1 provide sufficient conditions but not necessary conditions for the optimality of the obtained policy, in that (\ref{eqineq}) uses the inequality scaling technique.
	
	Now, we can achieve the goal (\ref{eqflipaction}) under the reward setting (\ref{eqr2}). Next, we present $Q$L for finding the optimal policy $\pi^*$.
	
	\vspace{-5pt}
	\subsection{Q-Learning for Finding Minimum Flipping Actions}
	In this subsection, we propose Algorithm 4 for finding minimum flipping actions for the reachability of BCNs under state-flipped control. 
	
	\begin{algorithm}[htb] 
		\caption*{ $\mathbf{Algorithm\ 4}$ Finding minimum flipping actions for reachability of  BCNs under state-flipped control using $Q$L} 
		\label{algKQL} 
		\begin{algorithmic}[1] 
			\REQUIRE  
			$\mathcal{M}_{0}$, $\mathcal{M}_{d}$, $B^*$, $\alpha_{t}\in (0,1]$, $\gamma=1$, $\epsilon\in[0,1)$, $N$, $T_{max}$\\
			\ENSURE  
			$\pi^*$
			\STATE Initialize $Q(x_t,a_t) \leftarrow 0,\forall x_t\in \mathcal{B}^n ,\forall a_t\in \mathcal{B}^{ m +|B^*|}$
			\FOR{ $ep = 0, 1,\dots, N-1 $ }
			\STATE $x_0 \leftarrow \operatorname{rand}(\mathcal{M}_{0})$
			\FOR{$t=0,1,\dots, T_{max}-1 $ and $x_t\notin\mathcal{M}_{d}$}
			\STATE $a_{t}\leftarrow\left\{
			\begin{aligned}
				&\arg \max _{a} Q(x_{t}, a),&&P=1-\epsilon\\
				&\operatorname{rand}(\mathcal{B}^{ m +|B|}),&&P=\epsilon
			\end{aligned}\right. $
			\STATE$Q(x_t,a_t)\leftarrow\alpha_{t}(r_{t+1}+\gamma\max\limits_{a}Q(x_{t+1}, a))+(1-\alpha_{t})Q(x_t,a_t)$
			\ENDFOR 
			\ENDFOR
			\RETURN $\pi^*(x_t)\leftarrow\arg \max\limits_{a}Q(x_t,a),\forall x_t\in  \mathbf{X}$ 
		\end{algorithmic}
	\end{algorithm}
	
	Next, we discuss the requirements for the parameters. In terms of $T_{max}$, it should be larger than the maximum length of the trajectory from $x_0$ to $\mathcal{M}_{d}$ without any cycle. Otherwise, the agent may not be able to reach the terminal state in $\mathcal{M}_{d}$ before the episode ends, which violates condition 4) of Lemma 1. \textcolor{black}{Furthermore, it should satisfy $\epsilon>0$, ensuring the exploration of the agent.} Hence, the policies during learning are non-deterministic, namely, all the policies can lead to a state $x\in\mathcal{M}_{d}$, which satisfies condition 4) of Lemma 1.
	
	Algorithm 4 is capable of finding $\pi^*$ for small-scale BCNs under state-flipped control. However, it becomes inapplicable when the values of $n$ and $m+|B|$ are too large since the $Q$-table is too large to be stored in computer memory. This motivates us to utilize small memory $Q$L in Section \uppercase\expandafter{\romannumeral4}-D.
	
	\vspace{-5pt}
	\subsection{Fast Small Memory Q-Learning for Finding Minimum Flipping Actions of Large-scale BCNs}
	In this section, we propose Algorithm 5 called the fast small memory $Q$L for finding the minimum flipping actions required to achieve reachability in large-scale BCNs under state-flipped control. Algorithm 5 differs from Algorithm 4 in two aspects.
	
	\begin{algorithm}[htb] 
		\caption*{ $\mathbf{Algorithm\ 5}$ Finding minimum flipping actions for reachability of large-scale BCNs under state-flipped control using fast small memory $Q$L} 
		\label{algKQL} 
		\begin{algorithmic}[1] 
			\REQUIRE 
			$\mathcal{M}_{0}$, $\mathcal{M}_{d}$, $B^*$, $\alpha_{t}\in (0,1]$, $\gamma=1$, $\epsilon\in[0,1)$, $N$, $T_{max}$, $w$, $\Delta w>0$\\
			\ENSURE  
			$\pi^*$
			\STATE Initialize $Q(x_t,a_t) \leftarrow 0,\forall x_t\in \mathcal{M}_{0}, \forall a_t\in \mathcal{B}^{ m +|B^*|}$
			\FOR{ $ep = 0, 1,\dots, N-1 $ }
			\STATE $x_0 \leftarrow \operatorname{rand}(\mathcal{M}_{0})$
			\IF {$w\leq\big|\{x_t|\ Q(x_t,.)$ is in the $Q$-table$\}\big|$}
			\STATE $w=w+\Delta w$
			\ENDIF
			\FOR{$t=0,1,\dots, T_{max}-1 $ and $x_t\notin\mathcal{M}_{d}$}
			\STATE $a_{t}\leftarrow\left\{
			\begin{aligned}
				&\arg \max _{a} Q(x_{t}, a),&&P=1-\epsilon\\
				&\operatorname{rand}(\mathcal{B}^{ m +|B|}),&&P=\epsilon
			\end{aligned}\right. $
			\IF {$Q(x_{t+1},.)$ is not in the $Q$-table}
			\STATE Add $Q(x_{t+1},.)=0$ to the $Q$-table
			\ENDIF
			\STATE$Q(x_t,a_t)\leftarrow\alpha_{t}(r_{t+1}+\gamma\max\limits_{a}Q(x_{t+1}, a))+(1-\alpha_{t})Q(x_t,a_t)$
			\ENDFOR 
			\ENDFOR
			\RETURN $\pi^*(x_t)\leftarrow\arg \max\limits_{a}Q(x_t,a),\forall x_t\in  \mathbf{X}$ 
		\end{algorithmic}
	\end{algorithm}
	
	Firstly, Algorithm 5 only records the action values of the visited states, as reflected in steps 1 and 9-11. This approach is similar to that employed in Algorithm 3. Secondly, in Algorithm 5, the parameter $w$ which affects $r_t$ changes from a constant to a variable that is dynamically adjusted according to $Q=\big|\{x_t|\ Q(x_t,.)$ is in the $Q$-table$\}\big|$, as shown in steps 4-6. As the agent explores more states, $\big|Q\big|$ increases and finally equals to \textcolor{black}{$|V\cup\mathcal{M}_{0}|$}. Here, we do not simply set $w=2^n+1-|\mathcal{M}_{d}|$ based on Corollary 1, in that for large-scale BCNs, the value of $2^n+1-|\mathcal{M}_{d}|$ is so large causing the weight for using fewer flipping actions to outweigh the importance of achieving reachability as soon as possible. This can lead to confusion for the agent. Specifically, at the beginning of the learning process, the agent receives large negative feedback when taking flipping actions, but less negative feedback when it fails to achieve reachability. The agent will only realize that taking flipping actions is worthwhile for achieving reachability when the accumulation of the negative feedback ``-1" for not realizing reachability exceeds that of ``$-w\times n_t$'' for flipping $n_t$ nodes. The larger the $w$, the longer the process will take. To address this issue, we introduce the scaling for $w$ to accelerate the process. The key concept behind this scaling method is to leverage the knowledge acquired by the agent while ensuring that the conditions in Theorem 4 are met to guarantee the optimality of the policy. We provide further explanation of this approach in Theorem 5.
	
	\noindent$\mathbf{Theorem\ 5}$.  For a BCN under state-flipped control (\ref{eqFlipBCN}), if it holds that $w>|Q|$, then the optimal policy obtained according to (\ref{eqr2}) satisfies the goal (\ref{eqflipaction}). 
	
	\noindent$\mathbf{Proof}$. 
	First, we prove that $\mathcal{M}_{d}$ is reachable from $\forall x_0\in \mathcal{M}_{0}$ only if, there exists $l\leq |V|$ such that $\mathcal{M}_{d}$ is $l$-step reachable from $\forall x_0\in \mathcal{M}_{0}$.
	Assume that $\mathcal{M}_{d}$ is reachable from $\mathcal{M}_{0}$. According to Definition 1, for any initial state $x_0 \in \mathcal{M}_{0}$, there exists a trajectory $x_0\rightarrow x_1
	...\rightarrow x_d$, where $x_d\in\mathcal{M}_{d}$. According to the definition of $V$, each state in the above trajectory is in $V$. Similar to the proof of Lemma 2, for any initial state $x_0 \in \mathcal{M}_{0}$, there exists a trajectory from $x_0$ to $\mathcal{M}_{d}$ whose length $l$ satisfies $l\leq |V|$.
	
	Next, we prove that $w>l$ after sufficient exploration by the agent. Since \textcolor{black}{$|Q|=|V\cup\mathcal{M}_{0}|$}, and $w>|Q|$, we have \textcolor{black}{$w>|V\cup\mathcal{M}_{0}|$}. From the proof in the previous paragraph, we know that $|V|\geq l$, therefore, we have $w>l$. By applying Theorem 4, we can guarantee the optimality of the policy. $\hfill\blacksquare$
	
	Thus, Algorithm 5 with the introduction of the variable $w$ still satisfies the goal defined by (\ref{eqflipaction}). Furthermore, it speeds up the learning process. Regarding the parameter settings of Algorithm 5, the initial $w$ can be set according to $Q=\big|\{x_t|\ Q(x_t,.)$ is in the $Q$-table$\}\big|$, where the $Q$-table is the one obtained from Algorithm 3.
	
	In summary, Algorithm 5 reduces the number of values to be stored from \textcolor{black}{$2^{n+m+|B^*|}$} to \textcolor{black}{$|V\cup\mathcal{M}_{0}|\times2^{m+|B^*|}$} by recording only the visited states. This approach is especially effective for solving large-scale BCN problems. Meanwhile, to accelerate the learning process, $w$ is set as a variable that will converge to a value that is larger than \textcolor{black}{$|V\cup\mathcal{M}_{0}|$}. 
	
	\vspace{-5pt}
	\section{Computational Complexity}
	
	The time complexities of Algorithms 1-5 involve selecting the optimal value from a pool of up to $2^{m+|B^*|}$ action-values per step, resulting in O($2^{m+|B^*|}$) complexity. 
		For Algorithms 1, 2, and 3, \textcolor{black}{the total number of steps} is expressed as iter$^{1,3}=\sum_{k=0}^{|B^*|-1}C_{|A|}^kNT_{\text{max}}+\sum_{k=0}^{C_{|A|}^{|B^*|}}N_{k}^{1,3}T_{\text{max}}$ and iter$^2=\sum_{k=0}^{|B^*|-1}C_{|A|}^kNT_{\text{max}}+\sum_{k=0}^{C_{|A|}^{|B^*|}}N_{k}^2T_{\text{max}}$ \textcolor{black}{respectively,} revealing pre- and post-flipping kernel discovery iterations. Introducing early termination based on reachability conditions in Algorithms 1-3 reduces the number of episodes for the $k^\text{th}$ flip set with $N_{k}^{1,3}\leq N$ and $N_{k}^2\leq N$, while Algorithm 2's efficiency ensures $N_{k}^{1,3}\leq N_{k}^2$. Algorithms 4 and 5 require $NT_{max}$ steps in total. Refer to Table \ref{complexity} for a comprehensive overview of the time complexities of each algorithm. 
		
		Concerning space complexity, Algorithms 1 and 4 store all action-values in a $Q$-table, while Algorithm 2 additionally manages tables for subsets of the current flip set. In contrast, Algorithms 3 and 5 retain action-values for solely visited states. Details on the final space complexities for each algorithm can be found in Table \ref{complexity}.

	\begin{table}
		\caption{\textcolor{black}{Time and space complexity of Algorithms 1 to 5}
		}
		\vspace{-10pt} 
		\setlength{\tabcolsep}{2.5pt}
		\renewcommand\arraystretch{1.1}
		\begin{center}
			\begin{tabular}{| p{15pt} |p{90pt} |  p{130pt}|}
				\hline
				& Time complexity&Space complexity\\
				\hline
				1 & $O$($2^{m+|B^*|}$iter$^1$)&$O$($2^{n+m+|B^*|}$)\\
				\hline
				2 & $O$($2^{m+|B^*|}$iter$^2$)&$O$($2^{n+m+|B^*|}(C_{|A|}^{|B^*|-1}+1)$)\\
				\hline
				3 & $O$($2^{m+|B^*|}$iter$^{1,3}$)&\textcolor{black}{$O$($2 ^ {m + | B^* |} $$| V\cup\mathcal{M}_{0} |$)}\\
				\hline
				4 & $O$($2^{m+|B^*|}$$NT_{max}$)&$O$($2^{n+m+|B^*|}$)\\
				\hline
				5 & $O$($2^{m+|B^*|}$$NT_{max}$)&\textcolor{black}{$O$($2 ^ {m + | B^*|} $$| V\cup\mathcal{M}_{0} |$)}\\
				\hline
			\end{tabular}
		\end{center}
		\label{complexity}
		\vspace{-15pt} 
	\end{table}
	
	\vspace{-5pt}
	\section{Simulation}
	In this section, the performance of the proposed algorithms is shown. Two examples are given, which are a small-scale BCN with 3 nodes and a large-scale one with 27 nodes. For each example, the convergence efficiency of different algorithms to check the reachability, and $\pi^*$ with minimal flipping actions are shown.
	
	\vspace{-5pt}
	\subsection{A Small-scale BCN}
	\noindent$\mathbf{Example\ 2}$. We consider a small-scale BCN  with the combinational flip set $A=\{1,2,3\}$, the dynamics of which is given as follows:
	\begin{equation}
		\left\{\begin{aligned}
			x_{1}(t+1)=&x_1(t)\wedge(x_2(t)\vee x_3(t))\vee \neg x_1(t)\wedge(x_2(t)\oplus \\
			&x_3(t)), \\
			x_{2}(t+1)=&x_1(t)\vee\neg x_1(t)\wedge(x_2(t)\vee x_3(t)), \\
			x_{3}(t+1)=&\neg(x_1(t)\wedge x_2(t)\wedge x_3(t)\wedge u(t))\wedge (x_3(t)\vee\\
			&(x_1(t)\vee\neg(x_2(t)\wedge u(t)))\wedge (\neg x_1(t)\vee(x_1(t)\\
			&\oplus x_2(t))\vee u(t))).
		\end{aligned}\right.
		\label{eqBCNsmall}
	\end{equation}
	In terms of the reachability, we set $\mathcal{M}_{d}=(0,0,1)$ and $\mathcal{M}_{0}=\mathcal{B}^n\setminus\mathcal{M}_{d}$. 
	
\begin{figure}
		\vspace{-10pt} 
		\centerline{\includegraphics[width=\columnwidth]{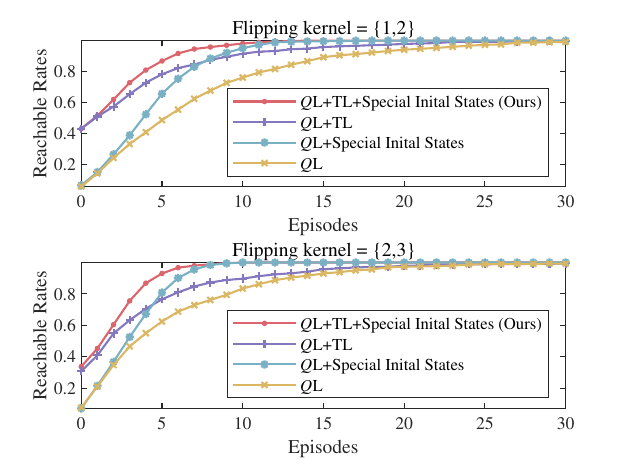}}
		\vspace{-5pt} 
		\caption{The convergence efficiency of Algorithms 1 and 2 for finding the flipping kernels of the small-scale system (\ref{eqBCNsmall})}
		\label{figSS}
		\vspace{-5pt} 
	\end{figure}

	To find the flipping kernels of system (\ref{eqBCNsmall}), Algorithms 1 and 2 are utilized. As mentioned in section \uppercase\expandafter{\romannumeral4}, Algorithm 2 has higher convergence efficiency compared with Algorithm 1. The result is verified in our simulation. To evaluate the convergence efficiency, the index $r_{ep}$ named reachable rate is defined as follows:
	\begin{equation}
		r_{ep}=\frac{nr_{ep}}{|\mathcal{M}_{0}|},
	\end{equation}
	where $nr_{ep}$ represents the number of $x_t\in\mathcal{M}_{0}$ that have been found to be reachable to $\mathcal{M}_{d}$ at the end of the $ep^{\text{th}}$ episode. For Algorithms 1 and 2, the flipping kernels utilized are $\{1,2\}$ and $\{2,3\}$, respectively. The effectiveness of Algorithms 1 and 2 is shown in Fig. \ref{figSS}, where the reachable rates are the average rates obtained from 100 independent experiments. In terms of $Q$L with TL, the initial reachable rate is higher since the agent utilizes prior knowledge. For $Q$L with special initial states, the reachable rate increases faster than conventional  $Q$L since the agent strategically selects the initial states. The best performance is achieved when both TL and special initial states are incorporated into $Q$L. 
	
\begin{figure}
		\vspace{-10pt} 
		\centerline{\includegraphics[width=\columnwidth]{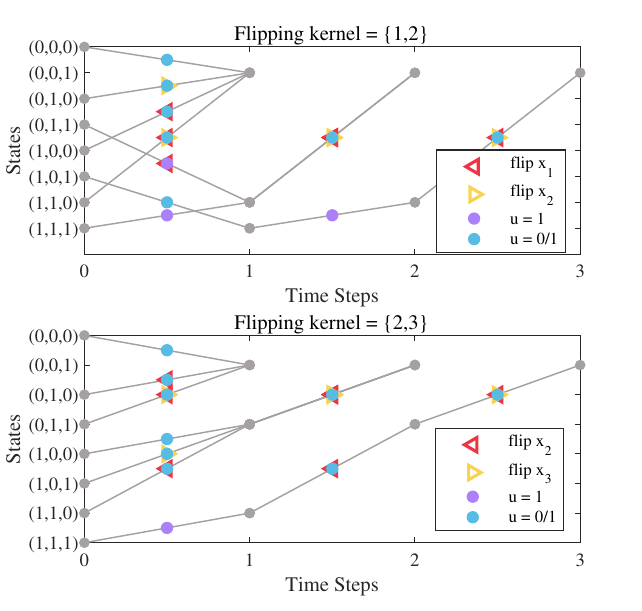}}
		\vspace{-5pt} 
		\caption{The optimal policy using the minimal flipping actions for reachability of system (\ref{eqBCNsmall}) obtained by Algorithm 4}
		\label{figSP}
		\vspace{-15pt} 
	\end{figure}

	Next, Algorithm 4 is employed to determine the minimum number of flipping actions required for reachability. The optimal policy obtained is shown in Fig. \ref{figSP}.

	\noindent$\mathbf{Remark\ 6.}$ \textcolor{black}{Emphasizing that the system model in this paper is solely for illustrative purposes, we stress that the agent has no prior knowledge of the model.}
	
	\noindent$\mathbf{Remark\ 7.}$ The flipping kernels obtained through Algorithms 1 and 2 have been compared with those obtained using the model-based method proposed by \cite{flip8}. Additionally, the policies depicted in Figs. \ref{figSP} and \ref{figPB} have been compared with other policies derived through enumeration. The results of these comparisons have verified that the flipping kernels and policies obtained are indeed optimal.
	
	\noindent$\mathbf{Remark\ 8.}$ \textcolor{black}{We utilize the method proposed by \cite{flip8} solely to validate the optimality of the flipping kernel, without directly comparing its computational complexity or convergence speed.} This is because the semi-tensor product method introduced by \cite{flip8} requires prior knowledge of the system model, whereas our approach is model-free. Therefore, it would be unfair to make a direct comparison between the two methods.
	
	\vspace{-5pt}
	\subsection{A Large-scale BCN}
	\noindent$\mathbf{Example\ 3}$. We consider a large-scale BCN with the combinational flip set $A=\{1,2,3,4,5,6\}$, the dynamics of which is given as follows:
	\begin{equation}
		\left\{\begin{aligned}
			x_{1}(t+1)=&x_1(t)\wedge(x_2(t)\vee x_3(t))\vee \neg x_1(t)\wedge(x_2(t)\oplus \\
			& x_3(t)), \\
			x_{2}(t+1)=&x_1(t)\vee\neg x_1(t)\wedge(x_2(t)\vee x_3(t)), \\
			x_{3}(t+1)=&\neg(x_1(t)\wedge x_2(t)\wedge x_3(t)\wedge u(t))\wedge (x_3(t)\vee\\
			&(x_1(t)\vee\neg(x_2(t)\wedge u(t)))\wedge (\neg x_1(t)\vee\\
			&(x_1(t)\oplus x_2(t))\vee u(t))),\\
			x_{3i-2}(t+1) = &	x_{3i-2}(t)\wedge(x_{3i-1}(t)\vee x_{3i}(t))\vee\neg x_{3i-2}(t)\\
			&\wedge(x_{3i-1}(t)\oplus x_{3i}(t)),\\
			x_{3i-1}(t+1) =& x_{3i-2}(t)\vee\neg x_{3i-2}(t)\wedge(x_{3i-1}(t)\vee x_{3i}(t)),\\
			x_{3i}(t+1) =& x_{3i}(t)\vee(\neg x_{3i-2}(t)\vee(x_{3i-2}(t)\oplus\\
			& x_{3i-1}(t))),\\
		\end{aligned}\right.
		\label{eqBCNlarge}
	\end{equation}
	where $i=2,3,...,9$. In terms of the reachability, we set $\mathcal{M}_{d}=(0,0,1,0,0,1,1,1,1,1,1,1,1,1,1,1,1,1,1,1,1,1,$ $1,1,1,1,1)$ and $\mathcal{M}_{0}=(a,0,0,1,0,0,1,0,0,1,0,0,1,0,0,$
	$1,0,0,1,0,0,1,0,0,1)$, where $a\in\{(0,0,0),(0,1,0),$
	$(0,1,1),$
	$(1,0,0),(1,0,1),(1,1,0),(1,1,1)\}$. 
	
	\begin{figure}
		\vspace{-10pt}
		\centerline{\includegraphics[width=\columnwidth]{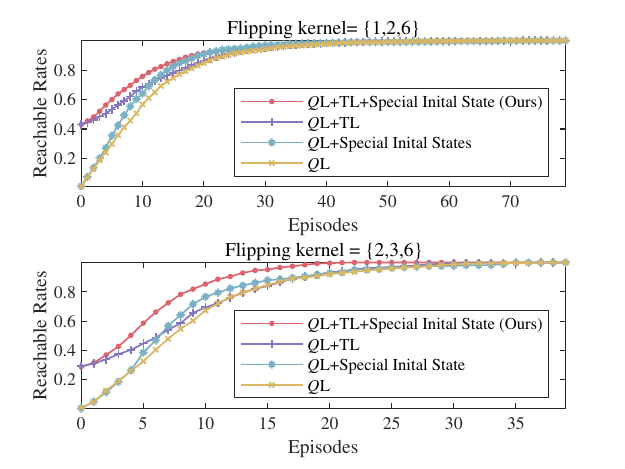}}
		\vspace{-5pt} 
		\caption{The convergence efficiency of Algorithm 3 and it with TL and special initial states for finding the flipping kernels of the large-scale system (\ref{eqBCNlarge})}
		\label{figSB}
		\vspace{-5pt} 
	\end{figure}
	
	\noindent$\mathbf{Remark\ 9.}$ System (\ref{eqBCNlarge}) consists of 9 small-scale BCNs. In Fig. \ref{figSB}, it can be observed that the number of flipping actions and time steps required \textcolor{black}{for system (\ref{eqBCNlarge})} to reach \textcolor{black}{state set $\mathcal{M}_{d}$} is relatively small. These two phenomena can be attributed to our intention of selecting an example that is easily verifiable for the optimal policy using the enumeration method. However, it should be noted that the proposed algorithms are applicable to all large-scale BCNs since the agent has no prior knowledge of the system model.

	\begin{figure}
		\vspace{-10pt}
		\centerline{\includegraphics[width=\columnwidth]{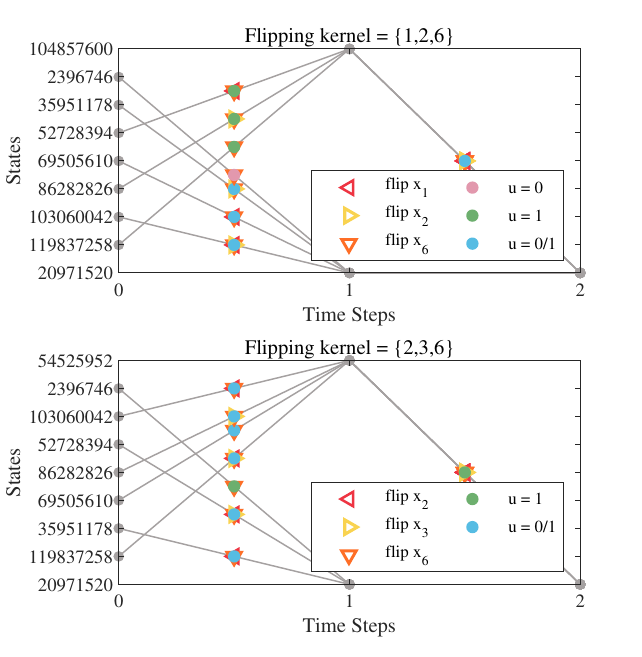}}
		\vspace{-5pt} 
		\caption{The optimal policy using minimal flipping actions to achieve the reachability for the large-scale system (\ref{eqBCNlarge}) obtained by Algorithm 5}
		\label{figPB}
		\vspace{-15pt} 
	\end{figure}

	To obtain the flipping kernels of the large-scale system (\ref{eqBCNlarge}), Algorithm 3 is employed along with the integration of TL and special initial states. The convergence efficiency of the \textcolor{black}{algorithms} is demonstrated in Fig. \ref{figSB}. According to Fig. \ref{figSB}, the combination of small memory $Q$L with both TL and special initial states yields the best performance.

	Subsequently, based on the flipping kernels $\{1,2,6\}$ and $\{2,3,6\}$ obtained from Algorithm 3, the optimal policy taking the minimal flipping actions to achieve the reachability is obtained using Algorithm 5. The resulting policy is displayed in Fig. \ref{figPB}. Notably, \textcolor{black}{in terms of reward setting,} if we set $w=2^{27}$ instead of utilizing a variable that changes according to $Q=\big|\{x_t|\ Q(x_t,.)$ is in the $Q$-table$\}\big|$, the policy cannot converge to the optimal one even with 10 times episodes. This demonstrates the effectiveness of Algorithm 5. In Fig. \ref{figPB}, we simplify the state in decimal form, where a state at time step $t$ is defined as $\sum_{i=1}^{27}2^{27-i}x_i(t)+1$.
	
	\vspace{-5pt}
	\subsection{Details}
	The parameters are listed as follows.
	\begin{itemize}[itemsep=1pt,topsep=1pt,parsep=1pt]
		\item In all algorithms, we set learning rate as $a_{t}=\min\{1,\frac{1}{(\beta ep)^\omega}\}$, which satisfies condition 1) in Lemma 1. 
		\item The greedy rate in all algorithms is specified as $1 - \frac{0.99ep}{N}$, gradually reducing from 1 to 0.01 as $ep$ increases. 
		\item We set $w=8$ for Algorithm 3 in Example 2 and $w=18$ with $\Delta w=20$ for Algorithm 5 in Example 3. These values of $w$ ensure the optimality of the obtained policy according to Corollary 1 and Theorem 4. 
		\item For Algorithms 1, 2, and 3, we set $\gamma=0.99$. 
	\end{itemize}
	$N$, $T_{max}$, $\beta$, and $\omega$ are selected based on the complexity of different examples and algorithms\textcolor{black}{, as detailed in Table \ref{tabDetail}.}
	
	\begin{table}[h]
		\caption{Parameter settings}
		\vspace{-10pt}
		\setlength{\tabcolsep}{2.5pt}
		\renewcommand\arraystretch{1.1}
		\begin{center}
			\begin{tabular}{| p{41pt} |p{48pt} |  p{31pt}| p{29pt}| p{29pt}| p{29pt}|}
				\hline
				Example&Algorithm&$N$&	$T_{max}$&	$\beta$	&$\omega$
				\\
				\hline
				2&1,2&100&	10&	1&	0.6\\
				\hline
				3&3&$10^4$	&$2^{27}$	&1&	0.6\\
				\hline
				2&4&$3\times10^4$&	100	&0.01&	0.85\\
				\hline
				3&5&$2\times10^5$&	$2^{27}$&	0.01&	0.85\\
				\hline
			\end{tabular}
		\end{center}
		\vspace{-15pt}
		\label{tabDetail}
	\end{table}

	\vspace{-5pt}
	\section{Conclusion}
	This paper presents $model$-$free$ reinforcement learning-based methods to obtain minimal state-flipped control for achieving reachability in BCNs, including large-scale ones. Two problems are addressed: 1) finding the flipping kernel, and 2) minimizing flipping actions based on the kernel.
	
	For problem 1) with small-scale BCNs, fast iterative  $Q$L is proposed. Reachability is determined using $Q$-values, while convergence is expedited through transfer learning and special initial states. Transfer learning migrates the $Q$-table based on the proven theorem that reachability preservation holds as the flip set size increases. Special initial states designate unknown reachability states to avoid redundant evaluation of known reachable states. For problem 1) with large-scale BCNs, we utilize small memory iterative $Q$L, which reduces memory usage by only recording visited action-values. Algorithm suitability is estimated via an upper bound on memory usage.
	
	For problem 2) with small-scale BCNs, $Q$L with BCN-characteristics-based rewards is presented. The rewards are designed based on the maximum length of reachable trajectories without cycles (an upper bound is given). This allows the minimization of flipping actions under terminal reachability constraints to be simplified as an unconstrained optimal control problem. For problem 2) with large-scale BCNs, fast small memory $Q$L with variable rewards is proposed. In this approach, the rewards are dynamically adjusted based on the maximum length of explored reachable trajectories without cycles \textcolor{black}{to enhance convergence efficiency}, and their optimality is proven.
	
	\textcolor{black}{Considering the critical value of reachability in controllability, our upcoming research will investigate minimum-cost state-flipped control
		for controllability of BCNs using reinforcement methods. Specifically, we will address two key challenges: 1) identifying the flipping kernel for controllability of BCNs, and 2) minimizing the required flipping actions to achieve controllability. While our approach can effectively address challenges 1 and 2 for small-scale BCNs with minor adjustments, adapting it for large-scale BCNs with simple modifications is presently impractical. This will be the primary focus of our future research.}
	
	\bibliography{reference}         
	
\end{document}